\documentclass{article}
\usepackage[utf8]{inputenc}
\usepackage{url}
\usepackage{geometry}\providecommand{\keywords}[1]
{
  \small	
  \textbf{\textit{Keywords---}} #1
}

\usepackage[justification=centering]{caption}
\usepackage{amsmath}
\usepackage[table, dvipsnames]{xcolor}
\usepackage{amssymb}

\usepackage{mathptmx}       
\usepackage{helvet}         
\usepackage{courier}        
\usepackage{type1cm}        
\usepackage{makeidx}         
\usepackage{graphicx}        
\usepackage{tabularx}
\usepackage{float}
\usepackage{subcaption}
\usepackage{multirow}
\usepackage{textcomp}
\usepackage[linesnumbered,ruled]{algorithm2e}
\usepackage{multicol}        
\usepackage[bottom]{footmisc}

\begin{document}

\title{Feature Extraction for Machine Learning-based Intrusion Detection in IoT Networks}
\author{Mohanad Sarhan*$^{1}$, Siamak Layeghy$^{1}$, Nour Moustafa$^{2}$,\\ 
Marcus Gallagher$^{1}$, Marius Portmann$^{1}$ \\
        \small $^{1}$University of Queensland, Brisbane QLD 4072, Australia \\
        \small $^{2}$University of New South Wales, Canberra ACT 2612, Australia\\
        \small *Corresponding Author: m.sarhan@uq.net.au\\
}
\date{}

\maketitle

\begin{abstract}
 A large number of network security breaches in IoT networks have demonstrated the unreliability of current Network Intrusion Detection Systems (NIDSs). Consequently, network interruptions and loss of sensitive data have occurred, which led to an active research area for improving NIDS technologies. In an analysis of related works, it was observed that most researchers aim to obtain better classification results by using a set of untried combinations of Feature Reduction (FR) and Machine Learning (ML) techniques on NIDS datasets. However, these datasets are different in feature sets, attack types, and network design. Therefore, this paper aims to discover whether these techniques can be generalised across various datasets. Six ML models are utilised: a Deep Feed Forward (DFF), Convolutional Neural Network (CNN), Recurrent Neural Network (RNN), Decision Tree (DT), Logistic Regression (LR), and Naive Bayes (NB). The accuracy of three Feature Extraction (FE) algorithms; Principal Component Analysis (PCA), Auto-encoder (AE), and Linear Discriminant Analysis (LDA), are evaluated using three benchmark datasets: UNSW-NB15, ToN-IoT and CSE-CIC-IDS2018. Although PCA and AE algorithms have been widely used, the determination of their optimal number of extracted dimensions has been overlooked. The results indicate that no clear FE method or ML model can achieve the best scores for all datasets. The optimal number of extracted dimensions has been identified for each dataset, and LDA degrades the performance of the ML models on two datasets. The variance is used to analyse the extracted dimensions of LDA and PCA. Finally, this paper concludes that the choice of datasets significantly alters the performance of the applied techniques. We believe that a universal (benchmark) feature set is needed to facilitate further advancement and progress of research in this field.

\end{abstract}




\section{Introduction}
\label{sec:introduction}

Cyber-security attacks and their associated risks have significantly increased since the rapid growth of the interconnected digital world \cite{stellios2018survey}, e.g., the Internet of Things (IoT) and Software-Defined Networks (SDN) \cite{sultana2019survey}. IoT is an ecosystem of interrelated digital devices and objects known as "things" \cite{khan2018iot}. They are embedded with sensors, computing chips and other technologies to collect and exchange data over the internet. IoT networks aim to increase the productivity of the hosting environment, such as industrial systems and "smart" buildings. IoT devices are growing significantly, with an expected number of 50 billion devices by the end of 2020 \cite{khan2018iot}. This growth has led to an increase in cyber attacks and the risks associated with them. Consequently, businesses and governments are proactively looking for new ways to protect their personal and organisational data stored on networked devices. Unfortunately, current security measures in IoT networks have proven unreliable against unprecedented attacks \cite{nawir2016internet}. For instance, in 2017, attackers compromised a casino's sensitive database through an IoT fish tank's thermometer. According to the Nozomi networks' report, new and modified IoT botnet attacks increased rapidly in the first half of 2020, with 57\% of IoT devices vulnerable to attacks \cite{pinto_2020}. According to the Symantec Internet Security Threat Report, more than 2.4 million new malware variants were created in 2018 \cite{symantec_2019}. That led to growing interest in improving the capabilities of NIDSs to detect unprecedented attacks. Therefore, new innovative approaches are required to enhance the attack detection performance of Network Intrusion Detection Systems (NIDSs). 

An NIDS is implemented in a network to analyse traffic flows to detect security threats and protect digital assets \cite{4656556}. It is designed to provide high cyber-security protection in operational infrastructures and aims to preserve the three principles of information systems security: confidentiality, integrity, and availability \cite{4656556}. Detecting cyber-attacks and threats have been the primary goal of NIDSs for a long time. There are two main types of NIDSs: Signature-based aims to match and compare the signatures of incoming traffic with a database of predetermined signatures of previously known attacks \cite{garcia-teodoro_diaz-verdejo_macia-fernandez_vazquez_2009}. Although they usually provide a high level of detection accuracy for precedented attacks, they fail to detect zero-day or modified threats that do not exist in the database. As attackers constantly change their techniques and strategies for conducting attacks to evade current security measures, NIDSs must be adaptive to evolving detection approaches. However, the current method for tuning signatures to keep up with changing attack vectors is unreliable. Anomaly-based NIDSs aim to overcome the limitations faced by signature NIDSs by using advanced statistical methods, which have enabled researchers to determine the behavioural patterns of network traffic. Various methods are used for anomaly detection, such as statistical-, knowledge- and Machine Learning (ML)-based techniques \cite{garcia-teodoro_diaz-verdejo_macia-fernandez_vazquez_2009}. Generally, they can achieve higher accuracy and Detection Rate (DR) levels for zero-day attacks, as they focus on matching attack patterns and behaviours rather than signatures \cite{amoli2016unsupervised}. However, anomaly NIDSs suffer from high False Alarm Rates (FARs) as they can identify any unique benign traffic that deviates from secure behaviour as an anomaly.

Current signature NIDSs have proven unreliable for detecting zero-day attack signatures \cite{hashemi2019towards} as they pass through IoT networks. This is due to the lack of known attack signatures in the system's database. To prevent these incidents from recurring, many techniques, including ML, have been developed and applied with some success. ML is an emerging technology with new capabilities to learn and extract harmful patterns from network traffic, which can be beneficial for detecting security threats \cite{sinclair1999application}. Deep Learning (DL) is an emerging branch of ML that has proven very successful in detecting sophisticated data patterns \cite{javaid2016deep}. Its models are inspired by biological neural systems in which a network of interconnected nodes transmits data signals. Each node contains a mathematical activation function that converts input to output. These models consist of hidden layers that can further extract complex patterns in network traffic. These patterns are learnt through network attack vectors, which can be obtained from various features transmitted through network traffic, such as packet count/size, protocols, services and flags. Each attack type has a different identifying pattern, known as a set of events that may compromise the security principles of networks if undetected.

Researchers have developed and applied various ML models, which are often combined with Feature Reduction (FR) algorithms to potentially improve their performance. Using a set of evaluation metrics, promising results for the detection capabilities of ML have been obtained, but these models are not yet reliable for real production IoT networks. The trend in this field is to outperform state-of-the-art results for a specific dataset rather than to gain insights into an ML-based NIDS application \cite{sommer2010outside}. Therefore, the extensive amount of academic research conducted outweighs the number of actual deployments in the real operational world. Although this could be due to the high cost of errors compared with those in other domains \cite{sommer2010outside}, it may also be that these techniques are unreliable in a real environment. This is because they are often evaluated on a single dataset consisting of a list of features that might not be feasible for collection or storage in a live IoT network feed. Moreover, due to the nature of ML, there is often room for improvement in its hyper-parameters when implemented on a specific dataset. Therefore, this paper aims to measure the generalisability of Feature Extraction (FE) algorithms and ML models combinations on different NIDS datasets.

In this paper, the effectiveness of three DL models in detecting attack vectors has been measured and compared with three Shallow Learning (SL) models, i.e., Deep Feed Forward (DFF), Convolution Neural Network (CNN), Recurrent Neural Network (RNN), Decision Trees (DT), Logistic Regression (LR) and Naive Bayes (NB). Three FE algorithms, namely, Principal Component Analysis (PCA), Linear Discriminant Analysis (LDA) and Auto-encoder (AE), have been explored, and their effects on three benchmark datasets, UNSW-NB15, ToN-IoT and CSE-CIC-IDS2018 have been studied. The results of the complete full dataset, without any FE algorithm applied, are also calculated for comparison. The extracted outputs of PCA and LDA are analysed by calculating their respective variance score. The optimal numbers of dimensions when applying the AE and PCA algorithms are found by experimenting with 1, 2, 3, 4, 5, 10, 20, and 30 dimensions. This paper is structured as follows; in Section \ref{related}, related works conducted in this field are explained. It is followed by a methodology section where the data processing, FE algorithms, and ML classifiers used and their architectures and parameters are mentioned. In Section \ref{results}, the datasets used and their importance in research are discussed, the evaluation metrics used are defined, and the results achieved are listed and explained. In summary, the key contributions of the paper are:

\begin{itemize}
    \item Experimental evaluation of 18 combinations of FE algorithms and ML classifiers across three NIDS datasets.
    
    \item Exploration of the number of feature dimensions and their impact on the classification performance.
    
    \item Analysis of feature variance and their correlation to the detection accuracy.

\end{itemize}

\section{Related works}
\label{related}

This section provides an overview of related papers and studies in this area. Due to the rapidly evolving nature of networks, new attack scenarios appear daily, and the age of a dataset is critical. As old datasets contain outdated patterns of benign and attack traffic, they are considered obsolete and have limited significance. Therefore, datasets released within the last five years are selected as they represent up-to-date network traffic. An updated version of CSE-CICIDS2017, known as CSE-CIC-IDS2018, was released publicly by the University of New Brunswick. Although the University of New South Wales released another dataset known as ToN-IoT in late 2019, limited papers that used it were found at the time of writing. Therefore, examining this dataset and its performance against well-known and widely used datasets is another contribution of this paper. Researchers have widely used the UNSW-NB15 dataset due to its various features and attack types. Papers in which the UNSW-NB15, ToN-IoT and CSE-CIC-IDS2018 datasets were used are analysed in the following paragraphs.

In \cite{azizjon_jumabek_kim_2020}, the authors implemented a CNN model and evaluated it on the UNSW-NB15 dataset. The CNN uses max-pooling, and a complete list of its hyper-parameters is provided. Experiments were conducted with different numbers of hidden layers and an addition of a Long Short Term Memory (LSTM) layer. The three-layer network performed best on the balanced and unbalanced datasets, achieving an accuracy of 85.86\% and 91.2\%, respectively, with the minority class oversampled to balance the label classes. The authors also compared three activation functions (sigmoid, relu, and tanh), with sigmoid obtaining the best accuracy of 91.2\%. Although they claimed to have built a reliable NIDS model, a DR of 96.17\% and FAR of 14\% are not ideal. They also did not evaluate their best model on various datasets to determine its stability or performance for different attack types or packet features. Khan et al. explored the five algorithms DT, RF, Gradient Boosting (GB), AdaBoost, and NB on the UNSW-NB15 dataset with an extra tree classifier for FE. The extracted features could have been heavily influenced by identifying features such as IPs and ports, which are biased towards attacking/victim nodes. The results showed that RF (98.60\%) achieved the best score, followed by AdaBoost (97.92\%) and DT (97.85\%). However, in terms of prediction times, DT performed the best with 0.75s, while RF and AdaBoost took 6.97s and 21.22s, respectively \cite{khan_sivaraman_honnavalli_2020}.

In \cite{larriva-novo_vega-barbas_villagra_sanz_rodrigo_2020}, the authors investigated various activation functions (relu, sigmoid, tanh, and softsign) and optimisers (adam, sgd, adagrad, nadam, adamax and RMSProp) with different numbers of nodes in the hidden layers. They aimed to find the optimal set of hyper-parameters for potential use in an NIDS. The experiment was conducted using DFF and LSTM architectures on the UNSW-NB15 dataset. There was no substantial improvement using LSTM rather than DFF, with the relu activation function outperforming the others. Most optimisers performed similarly well, except for SGD, which was less accurate. They claimed that their best setting for the hyper-parameters was using relu, adam, and a number of nodes following a configuration with the rule $0.75 \times \textrm{input} + \textrm{output}$. Their best accuracy results were 98.8\% for DFF and 98\% for LSTM. However, in the paper, neither the flow identifier features were dropped nor their best-claimed set of hyper-parameters evaluated on another dataset. In \cite{andalib2020novel}, the authors proposed an AE neural architecture consisting of LSTM and dense layers as an FE tool. The extracted output is then fed into an RF classifier to perform the attack detection. Three datasets, UNSW-NB15, ToN-IoT, and NSL-KDD, were used to evaluate the performance of the proposed methodology. The results indicate that the chosen classifier achieves higher detection performance without using compression methods. However, training time has been significantly reduced by using lower dimensions.

In \cite{zong_chow_susilo_2019}, the authors visually explored the effects of applying PCA and AE on the UNSW-NB15 and NSL-KDD datasets. They also experimented with different dimensions (ranging between 2 and 30) using the classifiers K Nearest Neighbour (KNN), DFF, and DT in a binary and multi-class classification scenario. The study found that AE performed better than PCA for KNN and DFF, but both were similar for DT. An optimal number of dimensions (20) was found for the UNSW-NB15 dataset but not for the NSL-KDD one. In \cite{tao_zhang_hu_hu_2019}, a CNN and an RNN model were designed to detect attacks in the CSE-CIC-IDS2018 dataset. The authors followed a supervised binary classification where CNN outperformed RNN in detecting each attack type. The authors have omitted some benign packets to balance attack and benign classes to improve classification performance. A significant increase in the performance was obtained in the detection of minority samples of attacks. Beloush et al. explored DT, NB, SVM and RF models on the UNSW-NB15 dataset. They have used accuracy as the defining metric where RF achieved 97.49\%, followed by a DT score of 95.82\%, and SVM and NB led to poor results. They applied no FR techniques, where the full dataset's features have been utilised. Training and testing times were also recorded, where NB achieved the fastest time \cite{belouch_el_hadaj_idhammad_2018}. 

In \cite{ferrag_maglaras_janicke_smith_2019}, the CSE-CIC-IDS2018 dataset has been utilised to explore seven different DL models, i.e., supervised (DFF, RNN and CNN) and unsupervised (restricted Boltzmann, DBN, deep Boltzmann machine and deep AE). The experiments also included a comparison of different learning rates and numbers of hidden nodes. However, any data pre-processing phase, including FR, was not mentioned. Moreover, the flow identifiers were not dropped, which would have caused a bias towards attacking victims' nodes or applications. All models performed similarly with slight variations in the DRs of their attack types. In terms of overall accuracy, CNN had the highest of 97.38\% when using 100 hidden nodes with 0.5 as the learning rate. Increasing the number of hidden nodes and learning rate improved the accuracy but also increased the training time. In \cite{qiao_blech_chen_2020}, the authors compared two FE techniques, namely, PCA and LDA, and proposed a linear discriminative PCA by feeding the discriminant information output from the LDA into the PCA. Although the ML model they used in their experiments was not identified, their method was evaluated on the UNSW-NB15 dataset. As their technique did not perform well for detecting fuzzers and exploits attacks, they decided to eliminate them from some of their results which are not ideal in a realistic network environment. Nevertheless, their results were still poor, with the best one for binary classification having a DR of 92.35\%. One of their stated future works is to determine the optimal number of principal components, i.e., the number of dimensions in a PCA. 

Most of the works found in the literature still adopted the negative habits addressed in \cite{sommer_paxson_2010}, with researchers aiming to create new FR methods and build new ML models to outperform the state-of-the-art results. However, due to the nature of the domain, researchers can always find a combination or variation that would result in slightly better numerical results. This can also be achieved by modifying any hyper-parameters used, which often have room for improvement when applied to a certain dataset. In most papers, experiments were conducted using a single dataset which questions the conclusion that their proposed techniques could be generalised across datasets. As each dataset contains its own private set of features, there are variations in the information presented. Consequently, these proposed techniques may have different performances, strongly influenced by the chosen dataset. The experimental issues mentioned above create a gap between the extensive academic research conducted on ML-based NIDSs and the actual deployments of ML-based NIDS in the operational world. However, compared with other applications, the same ML tools have been deployed in commercial scenarios with great success. We believe this is due to the high cost of errors in the NIDS domain, making it critical to design an optimal ML model before deployment. Therefore, as gaining insights into the ML-based NIDS application is crucial, this paper explores the performance of combinations of FE algorithms and ML models on different datasets. This will help determine if the best combination can be generalised for all chosen datasets. Also, although applying PCA and AE algorithms have been common in recent papers, finding the optimal number of dimensions to be used has been overlooked. The extracted dimensions of PCA and LDA are analysed by computing the variance and its correlation with the detection accuracy.


    
    
    
    




\section{Methodology}
\label{metho}

\begin{figure}[ht]
    \centering
    \includegraphics[width=7cm, height=8cm]{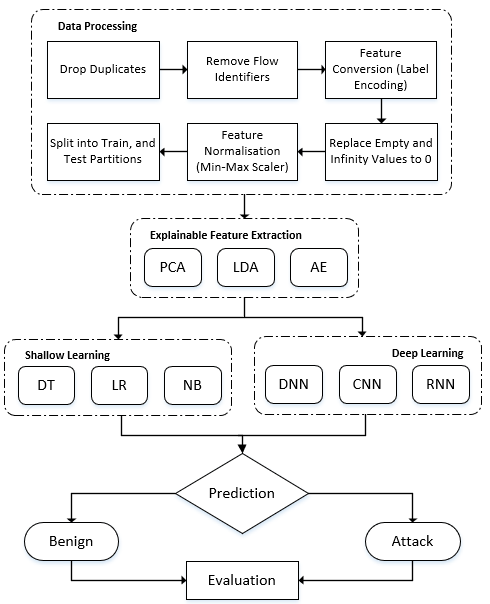}
    \caption{System architecture}
    \label{fig:system}
\end{figure}

This paper explores the effects of applying three FE techniques (PCA, LDA and AE) on three DL models (DFF, CNN and RNN) and three SL classifiers (DT, LR and NB). For PCA and AE, several dimensions (1,2,3,4,5,10,20 and 30) are selected to potentially find the optimal number. Three publicly released NIDS datasets that reflect modern network behaviour are utilised to conduct our experiments, with an overall representation provided in Fig.\ref{fig:system}. The datasets are processed for efficient FE and ML procedures. Then, the predictions made by the classifiers are collected, and certain evaluation metrics are statistically calculated. The Python programming language is used to design and conduct the experiments, and the TensorFlow and SciKitLearn libraries are used to build the DL and SL models, respectively.

\subsection{Data processing}
Data processing is an essential first step in enhancing the training process for ML models. All datasets are publicly available to download for research purposes. The duplicate samples (flows) are removed to reduce the storage size and avoid redundancy. Moreover, the flow identifiers, such as source/destination IP, ports and timestamps, are removed to prevent prediction bias towards the attacker's or a victim's end nodes/application. Then, the strings and non-numeric features are mapped to numerical values using a categorical encoding technique. These datasets contain features such as protocols and services, which are collected in their native string values, while the ML models are designed to operate efficiently with numerical values. There are two main techniques for encoding the features: one hot encoding and label encoding. The former transforms a feature into X categories by adding X number of features, using 1 to represent the presence of a category and 0 to represent its absence. However, this increases the number of dimensions of a dataset, which might affect the performance and efficiency of the ML models. Therefore, the label encoding technique maps each category to an integer. 

The nan, dash, and infinity values are replaced with 0 to generate a numerical-only dataset for use in the following steps. Any Boolean feature is replaced by 1 when it is true and 0 when it is false. Furthermore, the min-max feature scaling is applied to reduce complexity to bring all feature values between 0 and 1. It also allows all features to have equal weights, due to the nature of network traffic features, some values are larger than others, which can cause an ML model to pay more attention to them by assigning heavier weights. The min-max scaler computes all values of each feature by Eq.(\ref{mm}), where X\textsubscript{*} is the new feature value ranging from 0 to 1, X is the original feature value and X\textsubscript{max} and X\textsubscript{min} are the maximum and minimum values of the feature, respectively. The dataset is split into two portions for training and testing, and they are stratified based on the label features, which is essential due to the class imbalances of the datasets.

\begin{equation}X_*=\frac{X-X_{min}}{X_{max}-X_{min}}\label{mm}\end{equation}

\subsection{Feature extraction}
FE is the process of reducing the number of dimensions or features in a dataset. It aims to extract the valuable and relevant information spread among the raw input features and project it into a reduced number of features while minimising informational loss. The three FE algorithms used, PCA, LDA, and AE, are described in the following paragraphs. 

\begin{itemize}
    \item \textbf{Principal Component Analysis (PCA):} An unsupervised linear transformation algorithm that extracts features based on statistical procedures. It finds the eigenvectors with the highest eigenvalues in a covariance matrix and projects the dataset into a lower-dimensional space with a specified number of dimensions (features). These extracted features are an uncorrelated set called principal components. Although PCA is sensitive to outliers and missing values, it aims to reduce dimensionality without losing too much important or valuable information. The Singular Value Decomposition (SVD) solver is used in the PCA algorithm implemented in this paper. Different dimensions are explored to determine the effect of altering the input dimensions and find the optimal number of extracted features to use.

    \item \textbf{Linear Discriminant Analysis (LDA):} A supervised learning linear transformation algorithm that projects the features onto a straight line. It uses the class labels to maximise the distances between the mean of different classes (interclass) and minimise the distance between the mean of the same class (intraclass). It aims to produce features that are more distinguishable from each other. Similar to PCA, it aims to find linear combinations of features that help explain the dataset using a lower number of dimensions. However, unlike PCA, its number of extracted features needs to be equal to one less than its number of classes, which is one in our case, because there are two classes: attack and benign. LDA can also be utilised as a classification algorithm. However, in this paper, it is utilised as an FE technique where an SVD solver is implemented.

    \item \textbf{AutoEncoder (AE):} An artificial neural network designed to learn and rebuild feature representations. It contains two symmetrical components, an encoder and a decoder, with the former extracting a certain number of features from the dataset and the latter reconstructing them. When the number of nodes in the hidden layer is designed to be less than the number of input nodes, the model can compress the data. Therefore, during training, the model will learn to produce a lower-dimensional representation of the original input with the least loss of information. A dense AE architecture is used in these experiments because of the nature of the data. The number of nodes in the encoder block decreases in the order of 30, 20 and 10, and the decoder block increases in the reverse order. The number of nodes in the middle layer is set to the number of output dimensions required. All the layers consist of the relu activation function, adam optimiser and binary cross-entropy loss function.

\end{itemize}

\subsection{Machine learning}
ML is a subset of Artificial Intelligence (AI) that uses certain algorithms to learn and extract complex patterns from data. In the context of ML-based NIDS, ML models can learn harmful patterns from network traffic, which can be beneficial in the detection of security threats. DL is an emerging ML branch that is proven capable of detecting sophisticated data patterns. Its models are inspired by biological neural systems, in which a network of interconnected nodes transmits data signals. Building an ML model following a supervised classification method involves two processes: training and testing. During the first phase, the model is trained using labelled malicious and benign network packets from the training dataset to extract patterns and fit the corresponding model's parameters. Then, the testing phase evaluates the model's reliability by measuring its performance for classifying unseen attacks and benign traffic on the testing set of unlabelled network packets. These predictions are compared with the actual labels in the testing dataset to evaluate the model using certain metrics explained in Section \ref{emetrics}. 

The hyper-parameters used in the DL models are listed in Table \ref{NN}. All three datasets used in the experiments suffer from a class imbalance in terms of the frequency of benign and attack samples, which usually causes the model to predict the dominant class over the others. As the learning phase of an ML model is often biased towards the class with the majority of samples, the minority class may not be well fitted or trained in the final model \cite{prati_2018}. Due to the nature of the experiments, in two of the datasets, the minority class is an attack one, namely class 1, which is critical for the model to be able to detect and classify samples in that class. To deal with the datasets' imbalanced classes, weights are assigned to each class, with the minority having a "heavier" weight than the majority. Therefore, the model emphasises or gives priority to the former class in the training phase \cite{guo_yin_dong_yang_zhou_2008}. The classes’ weights are calculated using Eq.(\ref{c1}).

\begin{equation}W_{class}=\frac{Total Samples Count}{2 \times Class Samples Count}\label{c1}\end{equation}

\begin{itemize}
    \item \textbf{Deep Feed Forward (DFF):} A class of Multi-layer Perceptrons (MLPs) that is usually constructed of three or more hidden layers. In this model, the data is fed forward through the input layer and predictions are obtained on the outputs. Each layer consists of several nodes with weighted connections mapping the high-level features as input to the desired output. The weights are randomly initialised and then optimised in the learning phase through a process known as back-propagation. The input is a row (flow) of the CSV file fed into the input layer consisting of nodes equal to the number of input dimensions. Then, it is passed through three hidden layers consisting of 20 dense nodes, each having a relu activation function. The weight and biases are optimised using the adam algorithm with the binary cross-entropy loss function. Finally, due to our number of classes, the output layer is a single sigmoidal unit. The dropout rate of 0.2 is used to remove 20\% of the nodes' information to avoid over-fitting the training dataset. Fig.\ref{fig:DNN} presents the DFF architecture.

\begin{figure}[ht]
    \centering
    \includegraphics[width=8cm, height=4cm]{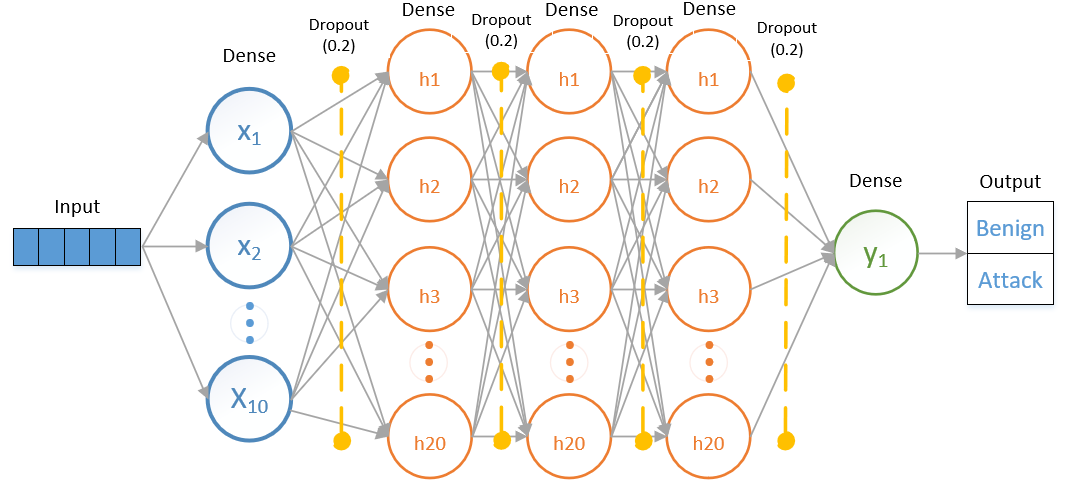}
    \caption{DFF model}
    \label{fig:DNN}
\end{figure}

\item \textbf{Convolution Neural Network (CNN):} An originally designed model to map images to outputs, which has proven to be effective when applied to any prediction scenario. Its hidden layers are typically convolutional and pooling ones, and a fully connected CNN includes an additional dense layer. Convolutional layers extract features with kernels from the input, and the pooling layers can enhance these features. The input is converted to a 2-dimensional shape to be compatible with the Conv1D layer. All layers have 20 filters, with kernel sizes of 3 in the input layer and 2 and 1 in the first and second hidden layers, respectively. All activation functions used in the convolutional layers are relu, and the average pooling size is 2 between each set of two convolutional layers. The input is passed to a dropout with a value of 0.2 and then to the final dense sigmoid classifier. Fig.\ref{fig:CNN} presents the mapping and pooling of the input by the convolutional layers until a prediction is made by the dense output layer. The hidden layers are removed for each input with less than 10 features, and its kernel size is reduced to 1.

\begin{figure}[ht]
    \centering
    \includegraphics[width=8cm, height=4cm]{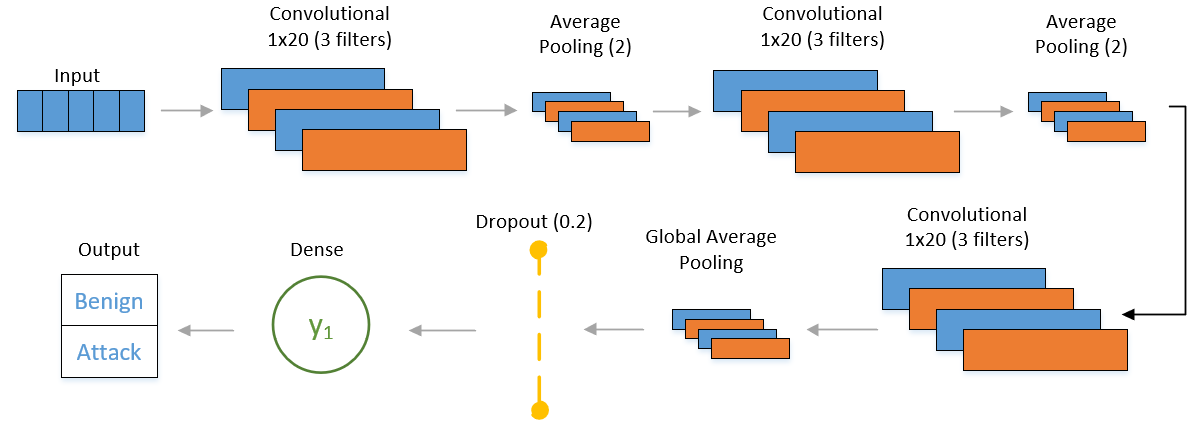}
    \caption{CNN model}
    \label{fig:CNN}
\end{figure}

\item \textbf{Recurrent Neural Network (RNN):} A model that can capture the sequential information present in input data while making predictions through an internal memory that stores a sequence of inputs, and it is successful in language-processing scenarios. Although there are various types of RNNs, LSTM is the most commonly used type of RNN. Each LSTM node contains three gates: forget, input, and output. The input is converted to a 3-dimensional shape to be compatible with the requirements of the LSTM layer. The number of nodes is equal to the number of input dimensions in the input layer. The input is then passed through a single hidden layer consisting of 10 nodes, with relu activation functions. Then, the weight and bias of each feature and layer are optimised using the adam algorithm based on the binary cross-entropy loss function. The output layer is a single sigmoidal output unit. The dropout rate of 0.2 is used to remove 20\% of the model's information to avoid over-fitting the training dataset. Fig.\ref{fig:LSTM} presents the mapping of an input to its output through LSTM layers.

\begin{figure}[ht]
    \centering
    \includegraphics[width=8cm, height=4cm]{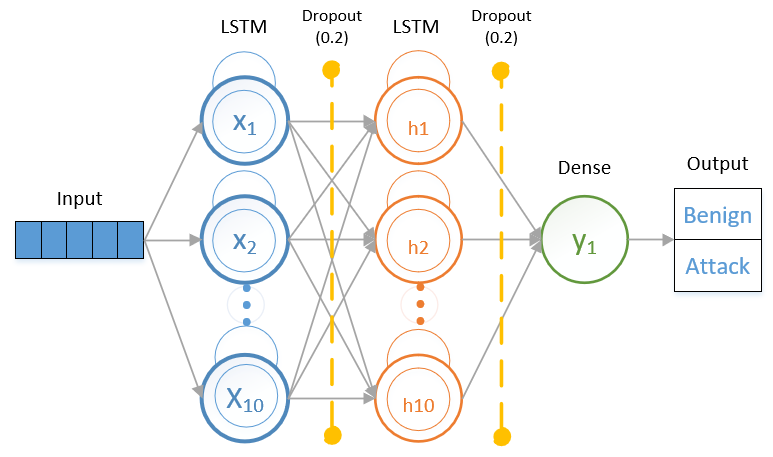}
    \caption{RNN model}
    \label{fig:LSTM}
\end{figure}

\begin{table}[ht] \footnotesize
\centering
\caption{DL hyper-parameters}
\label{NN}
\begin{tabularx}{\columnwidth}{|X|X|X|X|X|}\hline
\textbf{Parameter}                     & \textbf{DFF}        & \textbf{CNN Features $\geq$ 10} & \textbf{CNN Feature $<$ 10} & \textbf{RNN}        \\ \hline
\textbf{Layeras Type}                  & Dense               & Conv1D                             & Conv1D                             & LSTM                \\ \hline
\textbf{No. of Hidden Layer(s)}        & 3                   & 2                                  & N/A                                & 1                   \\ \hline
\textbf{Hidden Layer Neurons / Function} & 20 / Relu           & 20 / Relu                          & N/A                                & 10 / Relu           \\ \hline
\textbf{Output Layer Function}         & Sigmoid             & Sigmoid                            & Sigmoid                            & Sigmoid             \\ \hline
\textbf{Pooling Type / Size}             & N/A                 & Average / 2                        & Average / 2                        & N/A                 \\ \hline
\textbf{Optimi- sation}                  & Adam                & Adam                               & Adam                               & Adam                \\ \hline
\textbf{Loss}                          & Binary crossentropy & Binary crossentropy                & Binary crossentropy                & Binary crossentropy \\ \hline
\textbf{Dropout}                       & 0.2                 & 0.2                                & 0.2                                & 0.2                 \\ \hline
\end{tabularx}
\end{table}

\item \textbf{Logistic Regression (LR):} A linear classification model used for predictive analysis. It uses the logistic function, also known as the sigmoid function, to classify a binary output. It calculates the probability of being an output class between 0 and 1. It is easy to implement and requires few computational resources, but may not work well for non-linear scenarios. The lbfgs optimisation algorithm is selected with an l2 regularisation technique to specify the strategy for penalisation to avoid over-fitting. The tolerance value of the stopping criteria is set to 1e-4, the value of the regularisation strength to 1, and the maximum number of iterations to 100.

\item \textbf{Decision Trees (DT):} A model that follows a tree series in which each end node represents a high-level feature. The branches represent the outputs and the leaves represent the label classes. It uses a supervised learning method mainly for classification and regression purposes, aiming to map features and values to their desired outcome. It is widely used because it is easy to build and understand, but it can create an overcomplex tree that overfits the training data. The DT's Classification and Regression Trees (CART) algorithm is used due to its capability to construct binary trees using the input features \cite{ho1995random}. The Gini impurity function is selected to measure the quality of a split.


\item \textbf{Naive Bayes (NB):} A supervised algorithm that performs classification via the Bayes rule and models the class-conditional distribution of each feature independently. Although it is known to be efficient in terms of time consumption, it follows the "Naive" assumption of independence between each pair of input features. The Gaussian NB algorithm is chosen for its classification capabilities and retains the default value for variance smoothing of 1e-9. 
\end{itemize}

\subsection{Evaluation metrics}
\label{emetrics}
To evaluate the performances of the FE algorithms and ML models, the following evaluation metrics are used:
\begin{itemize}
    \item \textbf{TP :} True Positive is the number of correctly classified attack samples.
    \item \textbf{TN :} True Negative is the number of correctly classified benign samples.
    \item \textbf{FP :} False Positive is the number of misclassified attack samples.
    \item \textbf{FN :} False Negative is the number of misclassified benign samples.

    \item \textbf{Acc :} Accuracy is the number of correctly classified samples divided by the total number of samples: \begin{equation}ACC=\frac{TP+TN}{TP+TN+FP+FN} \label{acc}\end{equation}

    \item \textbf{DR :} Detection Rate, also known as recall, is the number of correctly classified attack samples divided by the total number of attack samples: \begin{equation}DR=\frac{TP}{TP + FN}\label{DR}\end{equation}
    
    \item \textbf{FAR :} False Alarm Rate is the number of incorrectly classified attack samples divided by the total number of benign samples: \begin{equation}FAR=\frac{FP}{FP+TN}\label{FAR}\end{equation}

    \item \textbf{F1 :} F1 Score is the harmonic mean of precision and DR:
    \begin{equation}Precision=\frac{TP}{TP + FP}\label{prec}\end{equation}
    
    \begin{equation}F1=\frac{2\times Precision\times DR}{Precision + DR}\label{f1}\end{equation}
    and
    
    \item \textbf{AUC :} Area Under the Curve is the area under the Receiver Operating Characteristics (ROC) curve that indicates the trade-off between the DR and FAR.
    
\end{itemize}

\begin{figure}[ht]
    \centering
    \includegraphics[width=4cm, height=4cm]{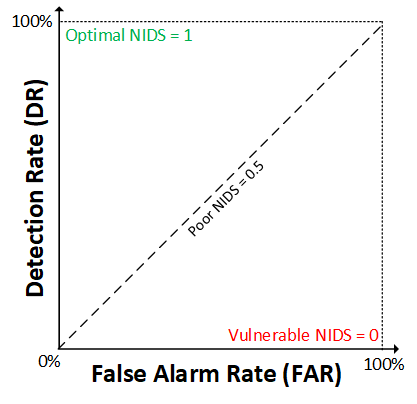}
    \caption{AUC}
    \label{fig:AUC}
\end{figure}\textbf{}

Most metrics are heavily affected by the imbalance of classes in the datasets. For example, a model can achieve a high accuracy or F1 score by predicting only the major class or having both a high DR and FAR, which makes it not ideal. Therefore, a single metric cannot be used to differentiate between models. The ROC considers both the DR and FAR by plotting them on the x- and y-axes, respectively, and then the AUC is calculated. This represents the trade-off between the two aspects and measures the performance of an NIDS for distinguishing between attack and benign flows. As shown in Fig.\ref{fig:AUC}, the ROC curve for an optimal NIDS is aimed toward the top left-hand corner of the graph with the highest possible AUC value of 1. On the other hand, an imperfect NIDS generates a graph of a diagonal line and has the lowest possible AUC value of 0.5.

\section{Results and discussion}
\label{results}

The following results are obtained from the testing sets using a stratified folding method of five-folds, and the mean results are calculated. In this section, the results for each dataset are initially discussed, and then all of them are considered for discussion. The early comparison of the models and FE algorithms is conducted using AUC as the comparison metric. For each dataset, the effects of applying the FE algorithms using different dimensions for each ML model are presented separately. Also, the best combination of an ML model and FE algorithm is selected to measure its performance for detecting each attack type statistically.

\subsection{Datasets}
Data selection is crucial for determining the reliability of ML models and the credibility of their evaluation phases. Obtaining labelled network data is challenging due to generation, privacy and security issues. Also, production networks do not generate labelled flows, which is mandatory when following a supervised learning methodology. Therefore, researchers have created publicly available benchmark datasets for training and evaluating ML models. They are generated through a virtual network testbed set up in a lab, where normal network traffic is mixed with synthetic attack traffic. The packets are then processed by extracting certain features using particular tools and procedures. An additional label feature is created to indicate whether a flow is malicious or benign. Each sample is defined by a network flow, with a flow considered a unidirectional data log between two end nodes where all the transmitted packets share specific characteristics such as IP addresses and port numbers. The following three datasets have been used:
\begin{itemize}
    \item \textbf{UNSW-NB15 :} A commonly adopted dataset released in 2015 by the Cyber Range Lab of the Australian Centre for Cyber Security (ACCS) \cite{moustafa-slay-2015}. The dataset originally contains 49 features extracted by Argus and Bro-IDS, now called Zeek tools. Although pre-selected training and testing datasets were created, the full dataset has been utilised. It has 2,218,761 (87.35\%) benign flows and 321,283 (12.65\%) attack ones, that is, 2,540,044 flows. Its flow identifier features are: \textit{id}, \textit{srcip}, \textit{dstip}, \textit{sport}, \textit{dport}, \textit{stime} and \textit{ltime}. The dataset contains non-integer features, such as \textit{proto}, \textit{service} and \textit{state}. The dataset contains nine attack types known as fuzzers, analysis, backdoor, Denial of Service (DoS), exploits, generic, reconnaissance, shellcode and worms.
    
    \item \textbf{ToN-IoT :} A recent heterogeneous dataset released in 2019 by ACCS \cite{fesz-dm97-19}. Its network traffic portion collected over an IoT ecosystem has been utilised, and it is made up of mainly attack samples with a ration of 796,380 (3.56\%) benign flows to 21,542,641 (96.44\%) attack ones, that is, 22,339,021 flows in total. It contains 44 original features extracted by Bro-IDS tool. The flow identifier features are named: \textit{ts}, \textit{src\_ip}, \textit{dst\_ip}, \textit{src\_port} and \textit{dst\_port}. It contains non-integer features, such as \textit{proto},  \textit{service} and \textit{conn\_state},  \textit{ssl\_version}, \textit{ssl\_cipher}, \textit{ssl\_subject}, \textit{ssl\_issuer}, \textit{dns\_query}, \textit{http\_method}, \textit{http\_version}, \textit{http\_resp\_mime\_types}, \textit{http\_orig\_mime\_types}, \textit{http\_uri}, \textit{http\_user\_agent}, \textit{weird\_addl} and \textit{weird\_name}. Its Boolean features include \textit{dns\_AA}, \textit{dns\_RD}, \textit{dns\_RA}, \textit{dns\_rejected}, \textit{ssl\_resumed}, \textit{ssl\_established} and \textit{weird\_notice}. The dataset includes multiple attack settings, such as backdoor, DoS, Distributed DoS (DDoS), injection, Man In The Middle (MITM), password, ransomware, scanning and Cross-Site Scripting (XSS). 
    
    \item \textbf{CSE-CIC-IDS2018 :} A dataset released by a collaborative project between the Communications Security Establishment (CSE) and Canadian Institute for Cybersecurity (CIC) in 2018 \cite{sharafaldin-habibi-lashkari-ghorbani-2018}. Their developed tool called CICFlowMeter-V3 was used to extract 75 network data features. The full dataset has been used, which has 13,484,708 (83.07\%) benign flows and 2,748,235 (16.93\%) attack ones, that is, 16,232,943 flows. Its flow identifier features are called \textit{Dst IP}, \textit{Flow ID}, \textit{Src IP}, \textit{Src Port}, \textit{Dst Port} and \textit{Timestamp}. Several attack settings were conducted, such as brute-force, bot, DoS, DDoS, infiltration, and web attacks.
    
\end{itemize}

\subsection{UNSW-NB15}

\begin{figure*}[ht]
\begin{subfigure}{.3\textwidth}
  \centering
  \includegraphics[width=5.5cm, height=3cm]{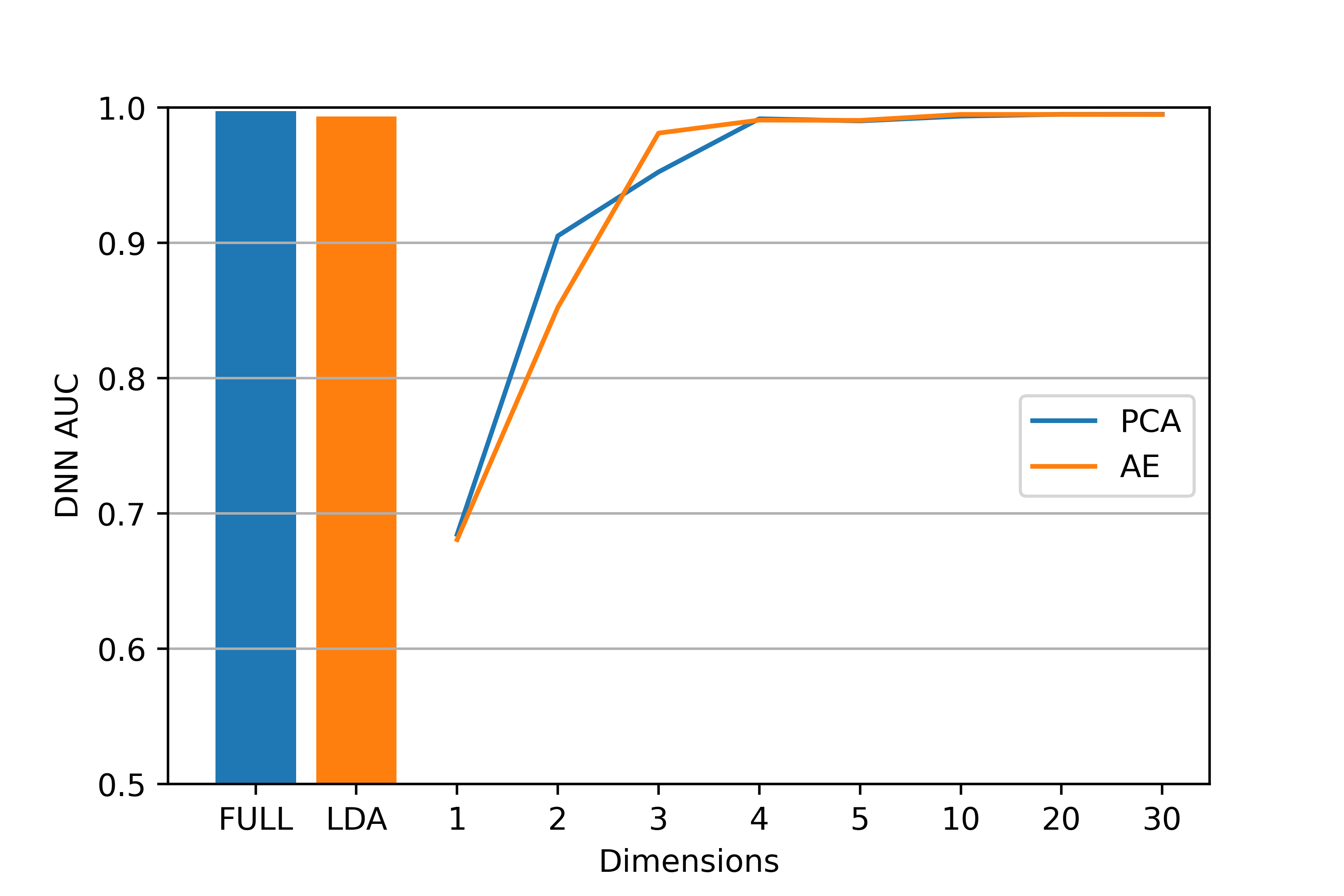}  
  \caption{DFF}
  \label{fig:dnnunsw}
\end{subfigure}
\hfill
\begin{subfigure}{.3\textwidth}
  \centering
  \includegraphics[width=5.5cm, height=3cm]{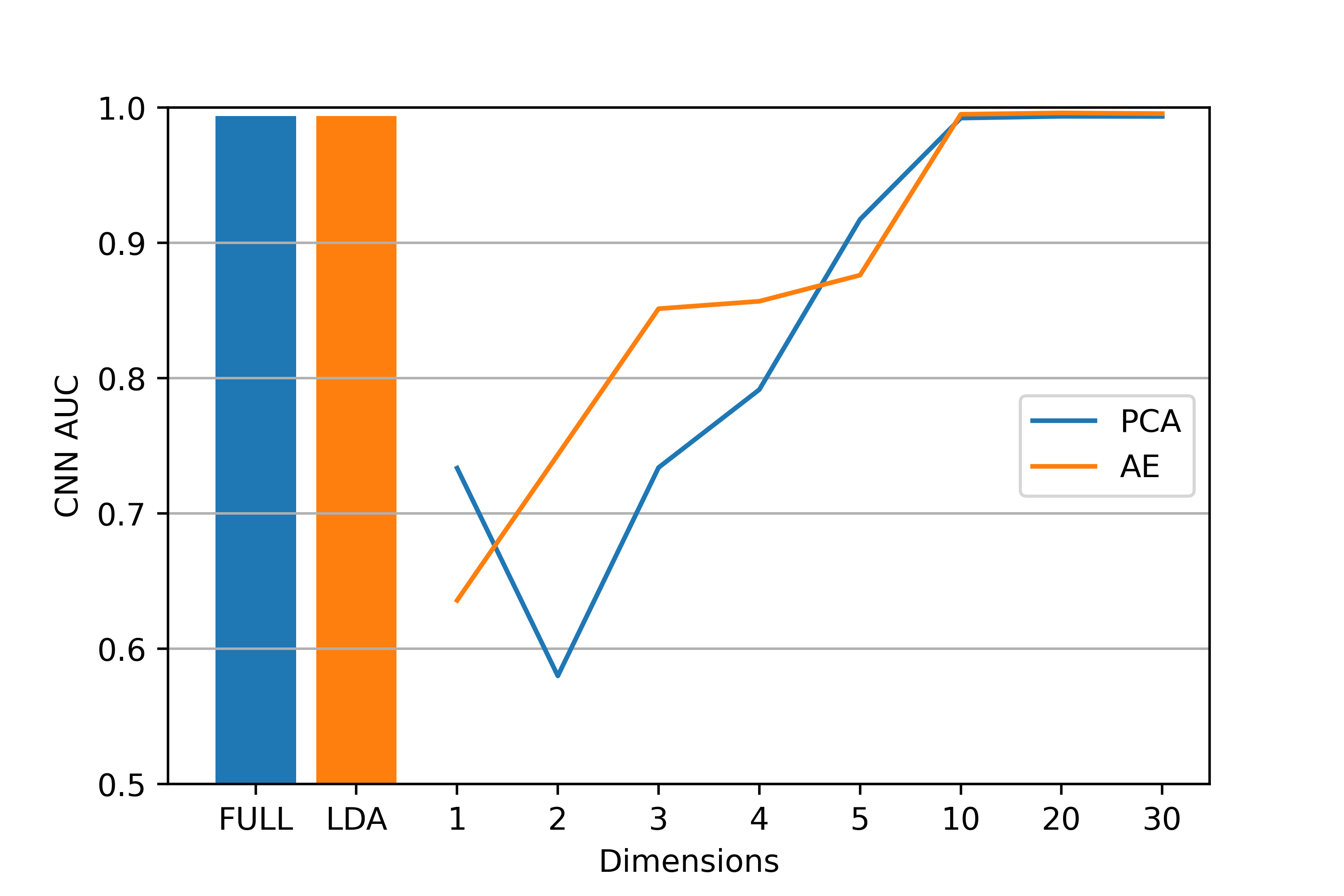}  
  \caption{CNN}
  \label{fig:cnnunsw}
\end{subfigure}
\hfill
\begin{subfigure}{.3\textwidth}
  \centering
  \includegraphics[width=5.5cm, height=3cm]{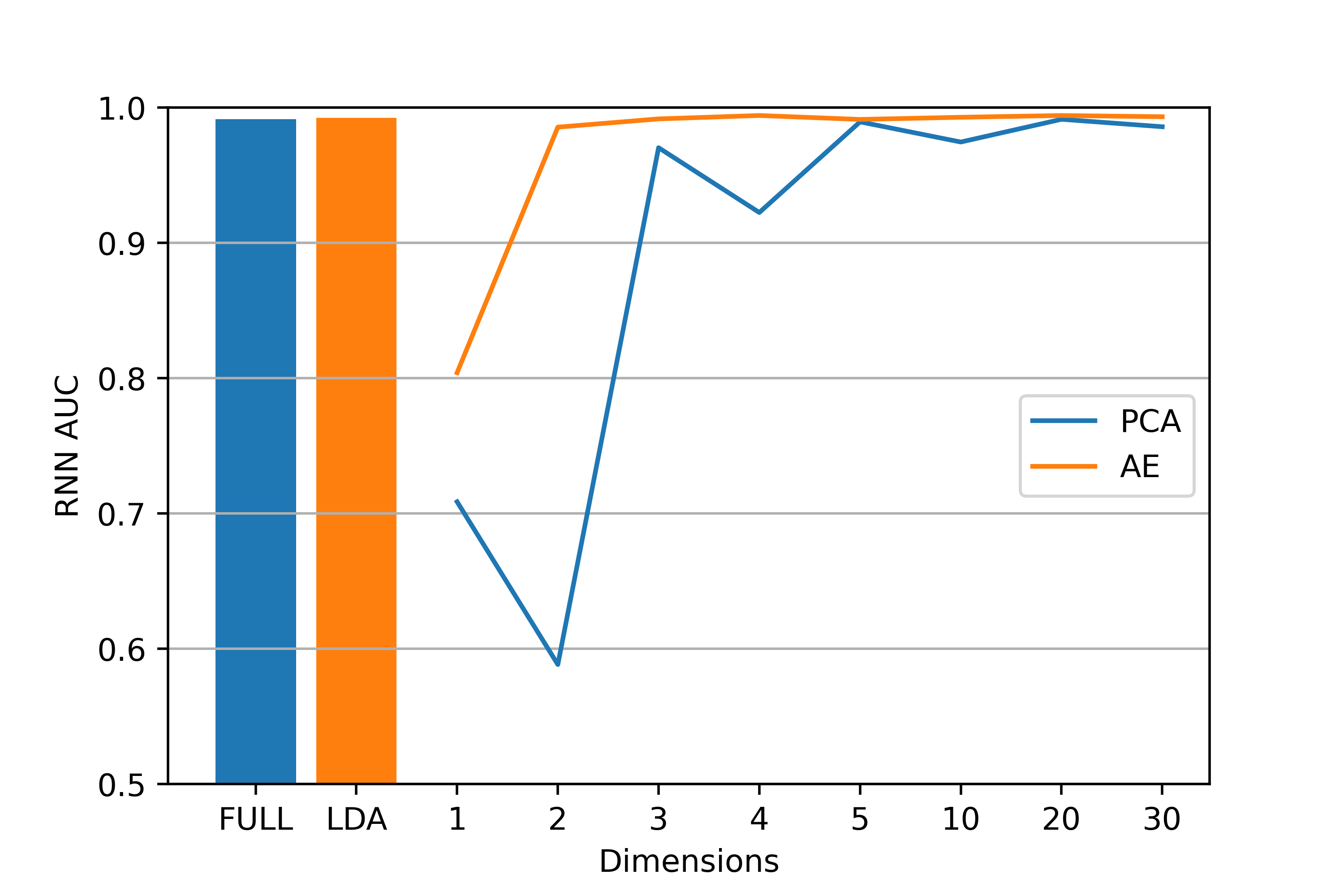}  
  \caption{RNN}
  \label{fig:rnnunsw}
\end{subfigure}
\hfill
\begin{subfigure}{.3\textwidth}
  \centering
  \includegraphics[width=5.5cm, height=3cm]{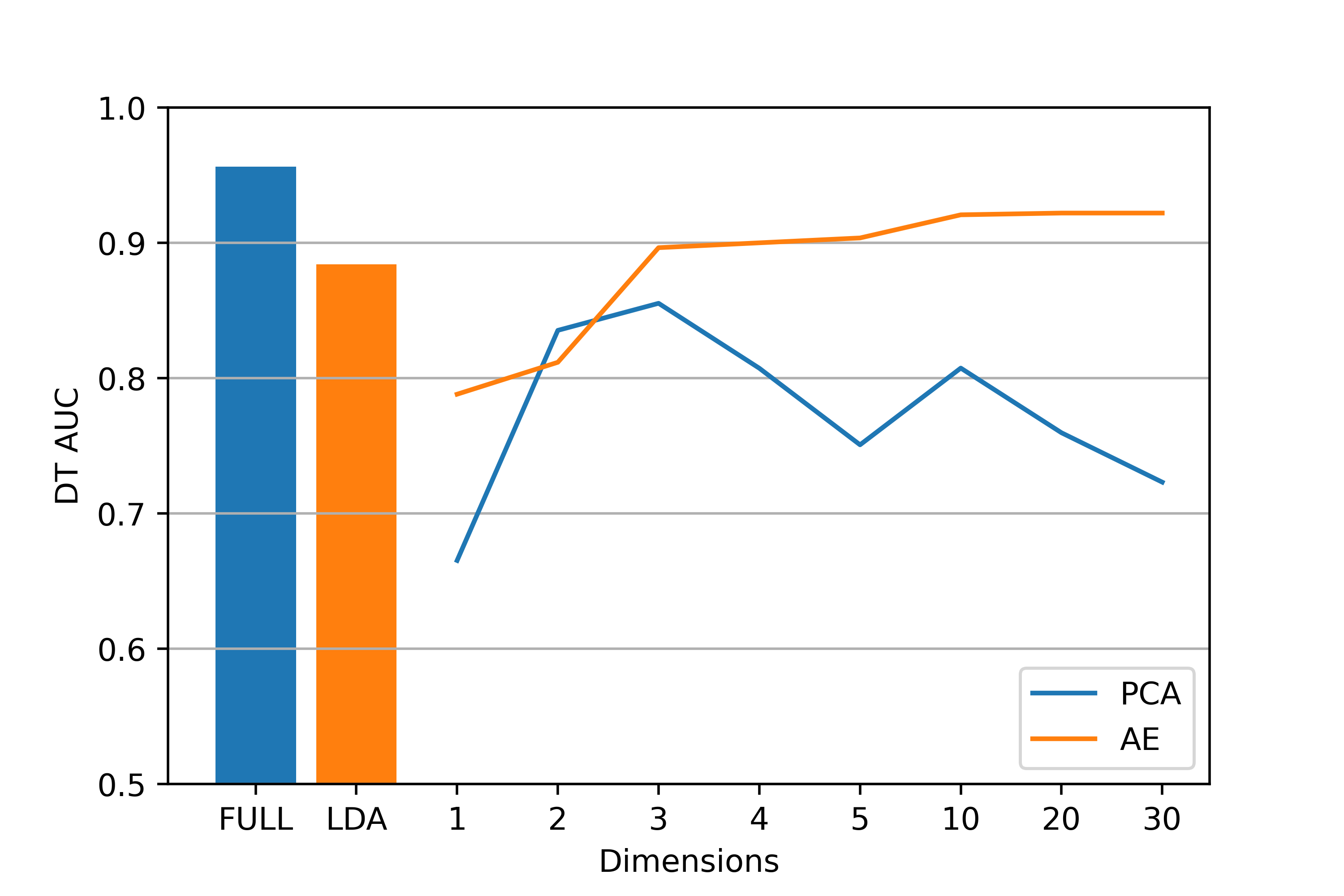}  
  \caption{DT}
  \label{fig:dtunsw}
\end{subfigure}
\hfill
\begin{subfigure}{.3\textwidth}
  \centering
  \includegraphics[width=5.5cm, height=3cm]{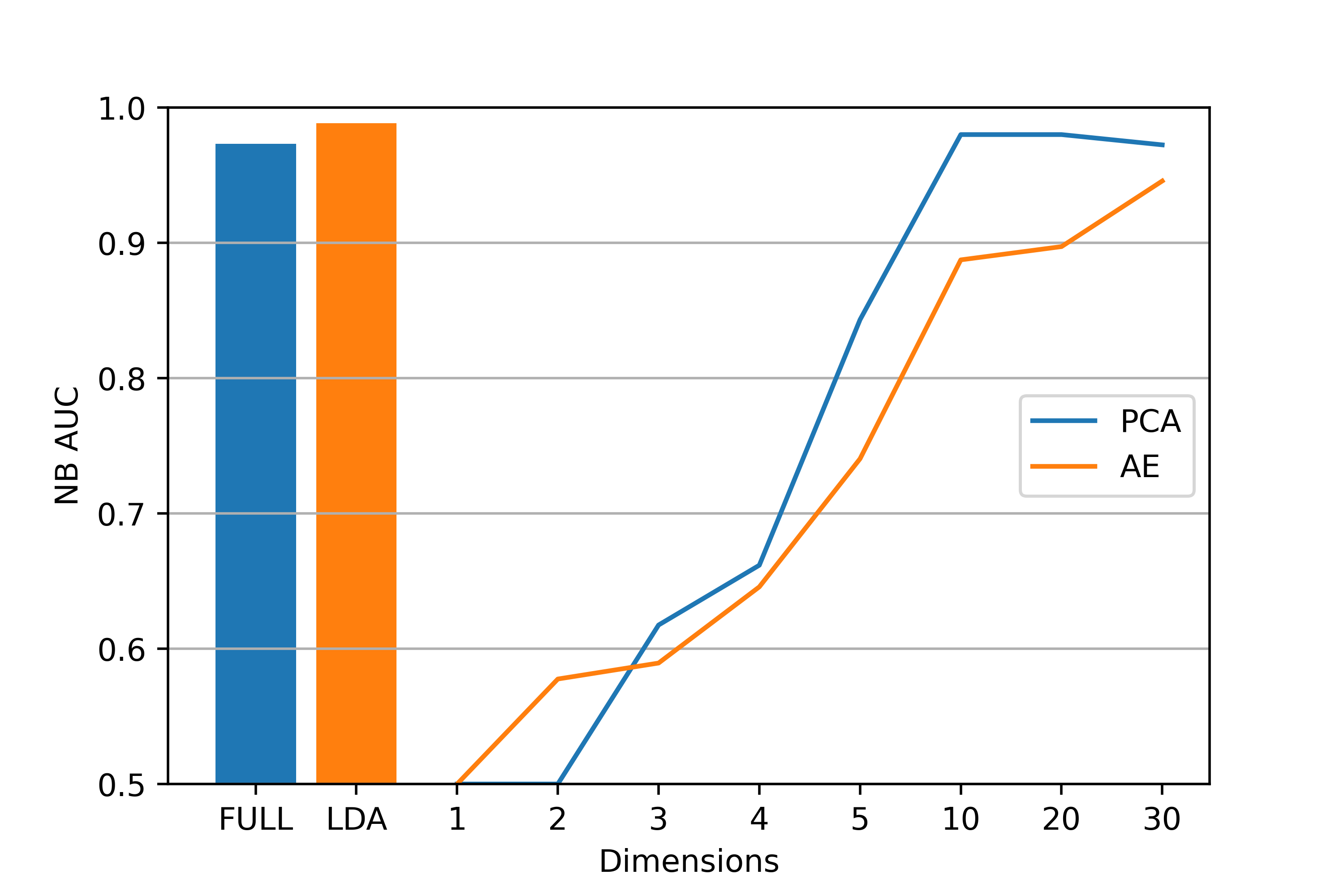}  
  \caption{NB}
  \label{fig:nbunsw}
\end{subfigure}
\hfill
\begin{subfigure}{.3\textwidth}
  \centering
  \includegraphics[width=5.5cm, height=3cm]{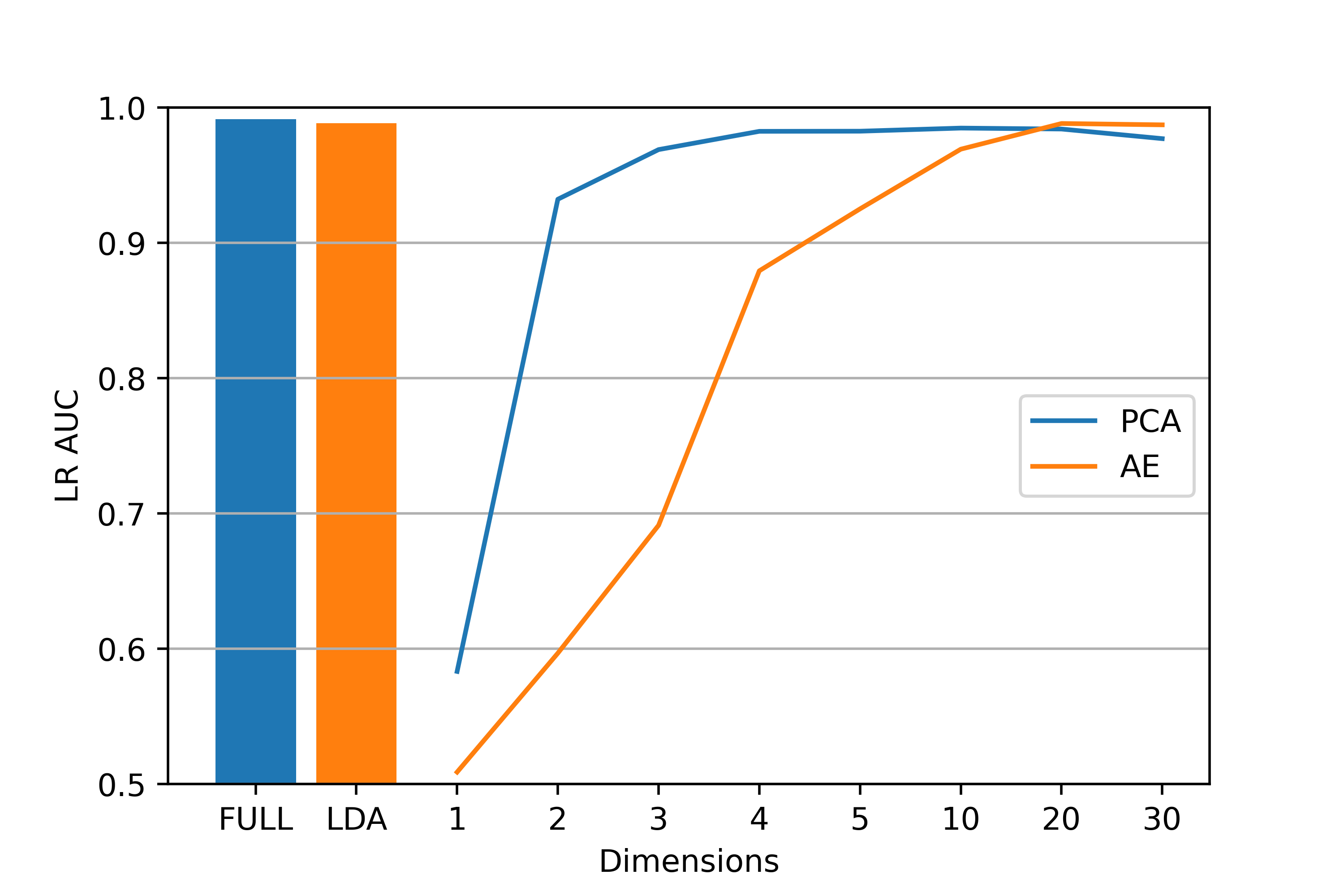}  
  \caption{LR}
  \label{fig:lrunsw}
\end{subfigure}
\caption{UNSW-NB15 results}
\label{fig:fig}
\end{figure*}

The results achieved on the UNSW-NB15 dataset by the ML models are similar in terms of their best performance, with DT obtaining the worst, as indicated in Fig.\ref{fig:fig}. The DFF and RNN models perform similarly where AE and PCA exponentially increase until dimension 3, when they start to fairly stabilise, while CNN requires 10 dimensions. AE improves the performances of CNN and RNN when using a low number of dimensions, whereas PCA with 2 dimensions reduces them significantly. All DL models perform equally well, achieving a generally higher AUC score than the SL classifiers. Although the NB and LR models achieve poor results when using a low number of dimensions for both AE and PCA, they improve rapidly with higher dimensions and obtain promising AUC scores. The performance of DT on the full dataset when using any of the FE algorithms is poor. PCA degrades the performance when the number of dimensions increases. LDA using a single dimension has achieved an excellent detection performance, similar to that achieved using the full dataset in most classifiers, where it achieved a higher score with NB but a lower score with DT.

\begin{table}[ht]\scriptsize
\centering
\caption{UNSW-NB15 classification metrics}\label{Metrics1}
\resizebox{\columnwidth}{!}{%
\begin{tabular}{|c|lllllll|}
\hline
\textbf{ML} &
  \multicolumn{1}{c}{\textbf{FE}} &
  \multicolumn{1}{c}{\textbf{DIM}} &
  \multicolumn{1}{c}{\textbf{ACC}} &
  \multicolumn{1}{c}{\textbf{F1}} &
  \multicolumn{1}{c}{\textbf{DR}} &
  \multicolumn{1}{c}{\textbf{FAR}} &
  \multicolumn{1}{c|}{\textbf{AUC}} \\ \hline
\multirow{4}{*}{DFF} & FULL & 40 & 98.33\% & 0.85 & 99.97\%   & 1.75\% & 0.9973 \\ \cline{2-8} 
                     & LDA & 1  & 98.34\% & 0.85 & 99.88\% & 1.74\% & 0.9935 \\ \cline{2-8} 
                     & PCA & 20 & 98.18\% & 0.84 & 99.87\% & 1.90\%  & 0.9954 \\ \cline{2-8} 
                     & AE  & 20 & 97.20\%  & 0.79 & 99.66\% & 2.92\% & 0.9949 \\ \hline
\multirow{4}{*}{CNN} & FULL & 40 & 98.22\% & 0.84 & 99.85\% & 1.86\% & 0.9938 \\ \cline{2-8} 
                     & LDA & 1  & 98.28\% & 0.85 & 99.89\% & 1.80\%  & 0.9937 \\ \cline{2-8} 
                     & PCA & 20 & 97.44\% & 0.80  & 99.31\% & 2.65\% & 0.9935 \\ \cline{2-8} 
                     & AE  & 20 & 98.16\% & 0.84 & 99.85\% & 1.92\% & 0.9960 \\ \hline
\multirow{4}{*}{RNN} & FULL & 40 & 98.12\% & 0.84 & 99.73\% & 1.97\% & 0.9915 \\ \cline{2-8} 
                     & LDA & 1  & 98.31\% & 0.85 & 99.88\% & 1.77\% & 0.9924 \\ \cline{2-8} 
                     & PCA & 20 & 97.89\% & 0.82 & 99.26\% & 2.18\% & 0.9913 \\ \cline{2-8} 
                     & AE  & 20 & 97.88\% & 0.83 & 99.88\% & 2.11\% & 0.9941 \\ \hline
\multirow{4}{*}{LR}  & FULL & 40 & 98.47\% & 0.86 & 99.88\% & 1.60\%  & 0.9914 \\ \cline{2-8} 
                     & LDA & 1  & 98.34\% & 0.84 & 99.41\% & 1.71\% & 0.9885 \\ \cline{2-8} 
                     & PCA & 10 & 98.13\% & 0.84 & 98.87\% & 1.91\% & 0.9848 \\ \cline{2-8} 
                     & AE  & 20 & 98.13\% & 0.84 & 99.59\% & 1.95\% & 0.9882 \\ \hline
\multirow{4}{*}{DT}  & FULL & 40 & 99.27\% & 0.92 & 91.58\% & 0.34\% & 0.9562 \\ \cline{2-8} 
                     & LDA & 1  & 97.86\% & 0.78 & 77.91\% & 1.10\%  & 0.8841 \\ \cline{2-8} 
                     & PCA & 3  & 97.41\% & 0.73 & 72.37\% & 1.31\% & 0.8553 \\ \cline{2-8} 
                     & AE  & 20 & 98.67\% & 0.86 & 85.15\% & 0.65\% & 0.9226 \\ \hline
\multirow{4}{*}{NB}  & FULL & 40 & 95.94\% & 0.70  & 98.82\% & 4.20\%  & 0.9731 \\ \cline{2-8} 
                     & LDA & 1  & 98.34\% & 0.85 & 99.39\% & 1.71\% & 0.9884 \\ \cline{2-8} 
                     & PCA & 20 & 97.47\% & 0.79 & 99.74\% & 2.65\% & 0.9854 \\ \cline{2-8} 
                     & AE  & 30 & 97.02\% & 0.75 & 91.87\% & 2.72\% & 0.9457 \\ \hline
\end{tabular}
}
\end{table}

The best results obtained by each FE algorithm using each ML model are listed in Table \ref{Metrics1}. When using the AE technique, the CNN performs best among other classifiers, with a high AUC score of 0.9960. AE improves the performance of CNN and RNN models compared to the other FE techniques. LR and DT achieve their best performances when applied to the full dataset without using any FE algorithm. LDA and PCA significantly degrade the performance of DT by decreasing the DR of attacks. However, they improve the NB classifier DR and lower its FAR. Interestingly, LDA performs better than PCA in all ML models except DFF, indicating an extreme correlation between one of the dataset's features and labels. Overall, the optimal number of dimensions for PCA and AE appears to be 20, which matches the findings in \cite{zong_chow_susilo_2019}. In Table \ref{UNSW-NB15 Multi}, the best-performing ML model has been applied, which is CNN, when using the AE technique with 20 dimensions to measure the DR of each attack type. It is confirmed that each attack type in the test dataset is almost fully detected, with backdoor and DoS attacks obtaining the lowest DRs.

\begin{table}[ht]\scriptsize
\centering
\caption{UNSW-NB15 attacks detection}
\label{UNSW-NB15 Multi}
\begin{tabular}{|c|c|c|c|}
\hline
\textbf{Attack Type} & \textbf{Actual} & \textbf{Predicted} & \textbf{DR} \\ \hline
Analysis          & 2185            & 2182               & 99.87\%        \\
Backdoor          & 1984            & 1966               & 99.10\%            \\
DoS               & 5665            & 5621              & 99.23\%               \\
Exploits          & 27599           & 27532              & 99.76\%            \\
Fuzzers           & 21795           & 21780              & 99.93\%         \\
Generic           & 25378           & 25355              & 99.91\%         \\
Reconnaissance    & 13357           & 13342              & 99.89\%      \\
Shellcode         & 1511            & 1511               & 100\%               \\
Worms             & 171             & 171                & 100\%              \\ \hline
\end{tabular}
\end{table}

\subsection{ToN-IoT}

\begin{figure*}[ht]
\begin{subfigure}{.3\textwidth}
  \centering
  \includegraphics[width=5.5cm, height=3cm]{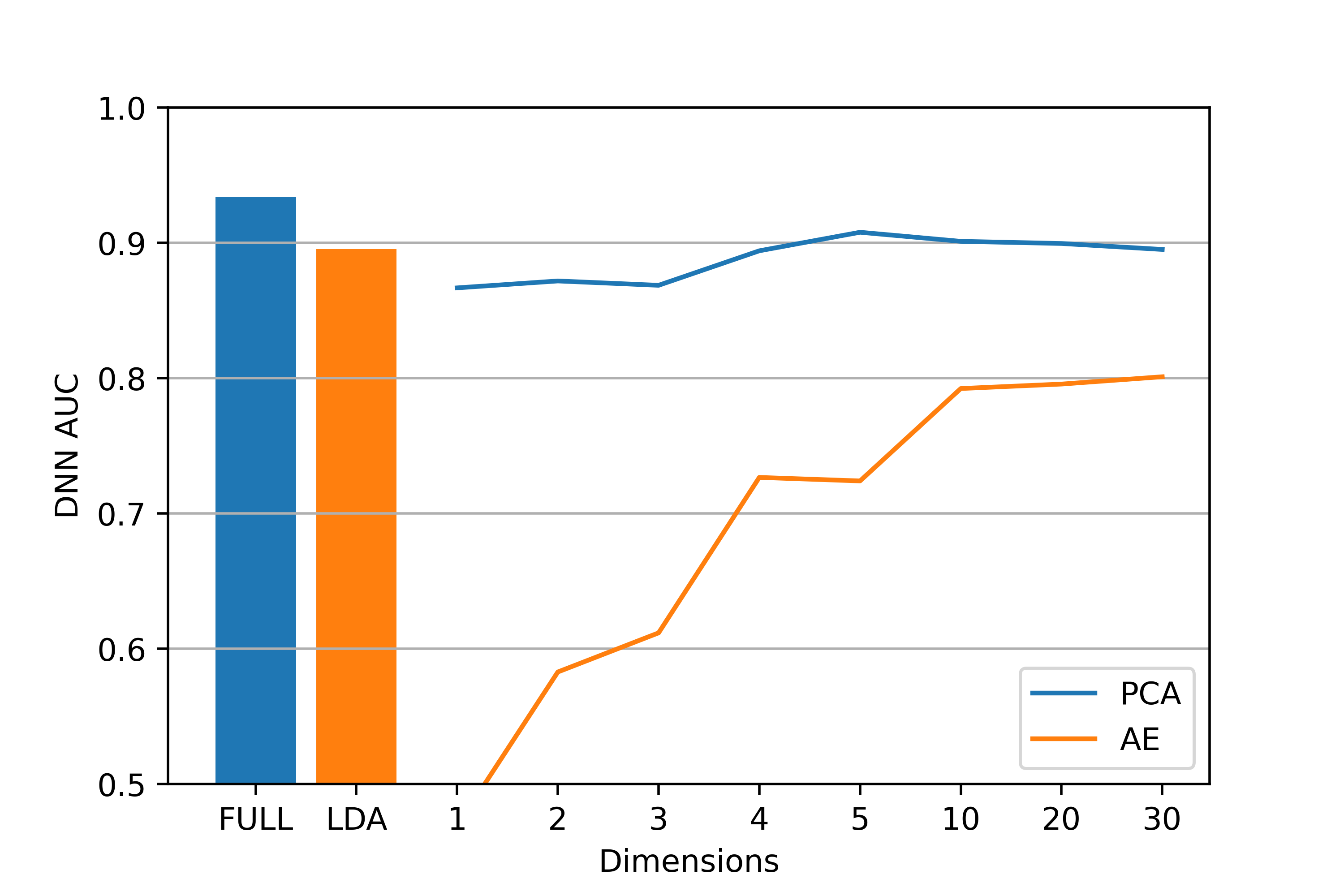}  
  \caption{DFF}
  \label{fig:dnnton}
\end{subfigure}
\hfill
\begin{subfigure}{.3\textwidth}
  \centering
  \includegraphics[width=5.5cm, height=3cm]{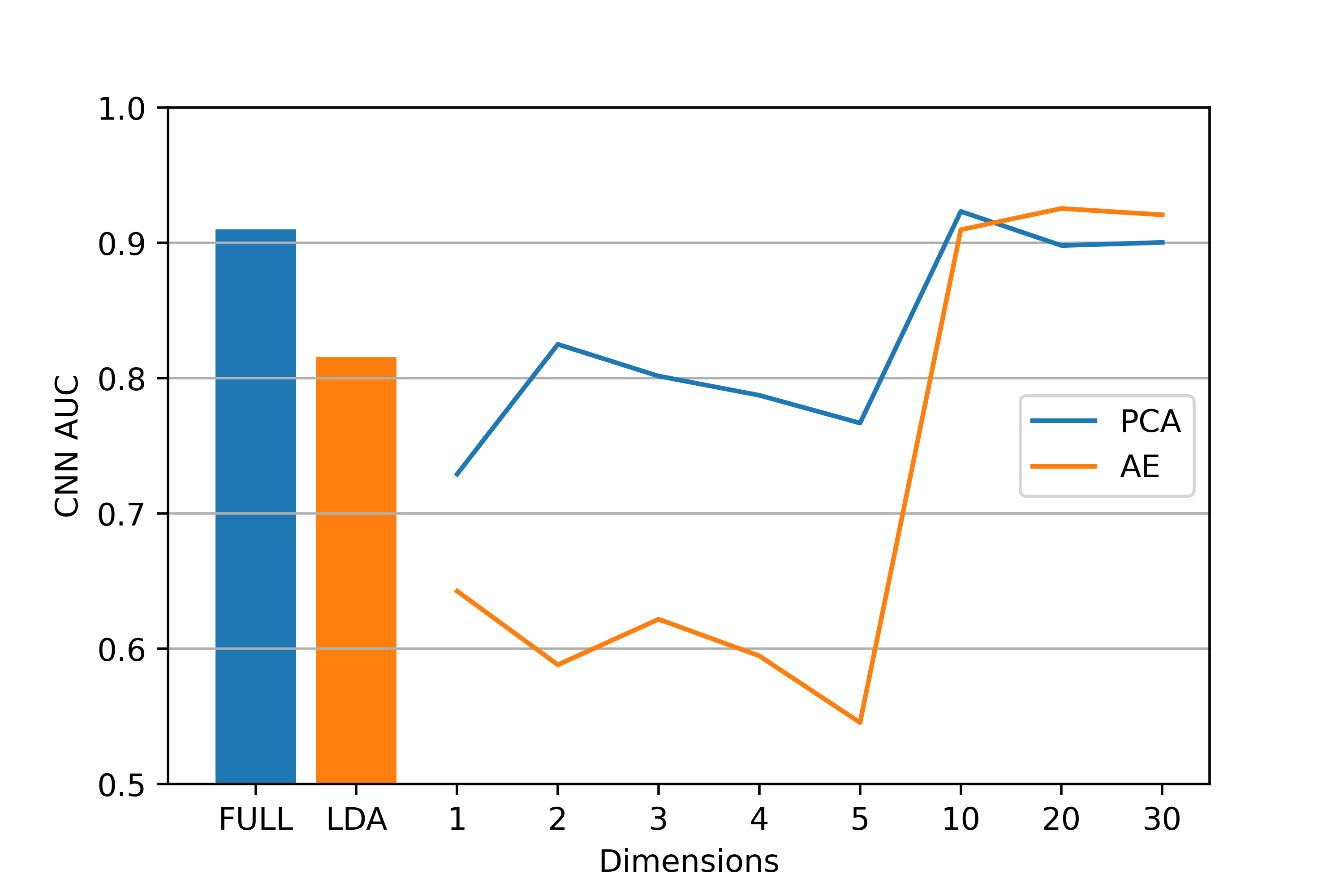}  
  \caption{CNN}
  \label{fig:cnnton}
\end{subfigure}
\hfill
\begin{subfigure}{.3\textwidth}
  \centering
  \includegraphics[width=5.5cm, height=3cm]{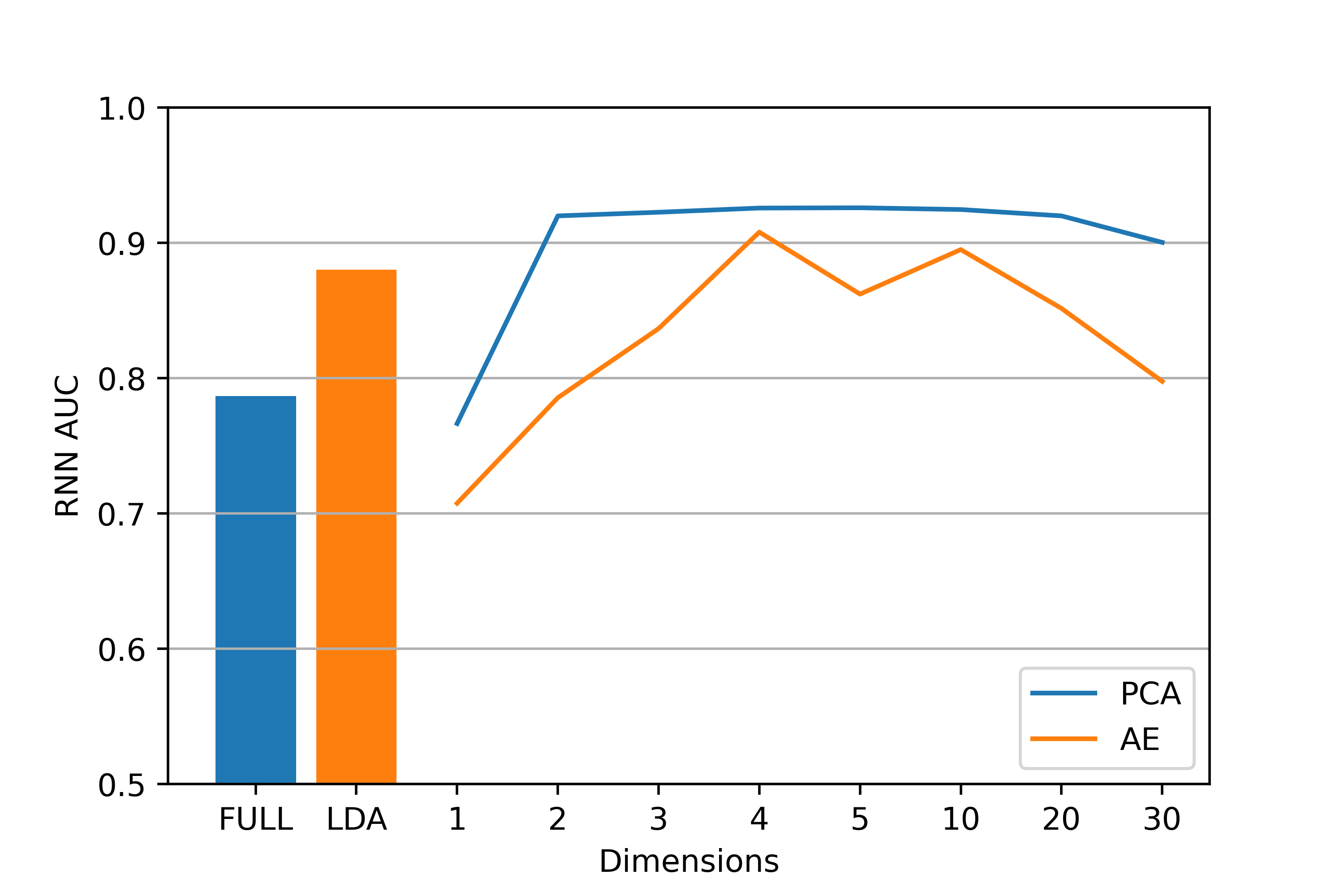}  
  \caption{RNN}
  \label{fig:rnnton}
\end{subfigure}
\hfill
\begin{subfigure}{.3\textwidth}
  \centering
  \includegraphics[width=5.5cm, height=3cm]{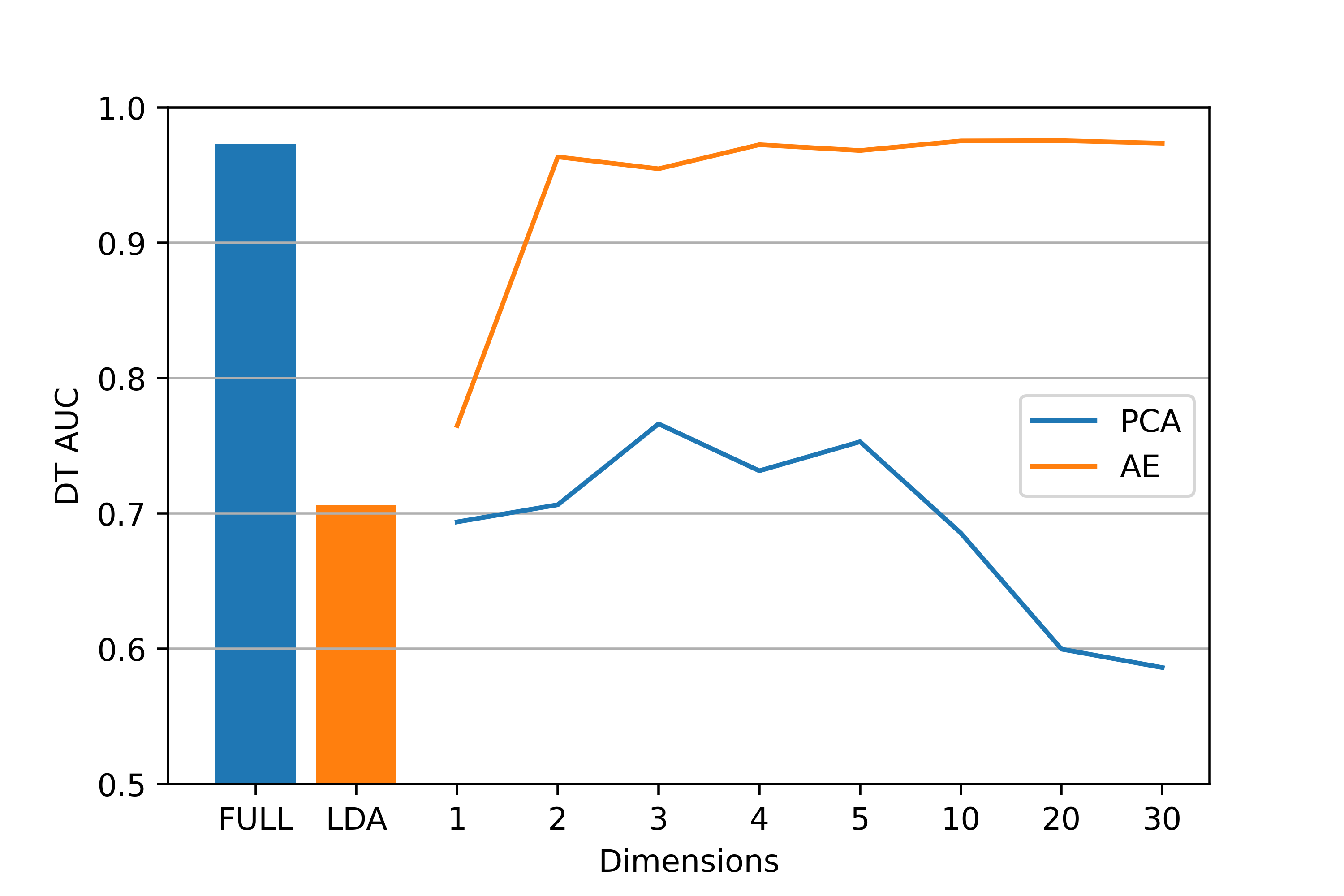}  
  \caption{DT}
  \label{fig:dtton}
\end{subfigure}
\hfill
\begin{subfigure}{.3\textwidth}
  \centering
  \includegraphics[width=5.5cm, height=3cm]{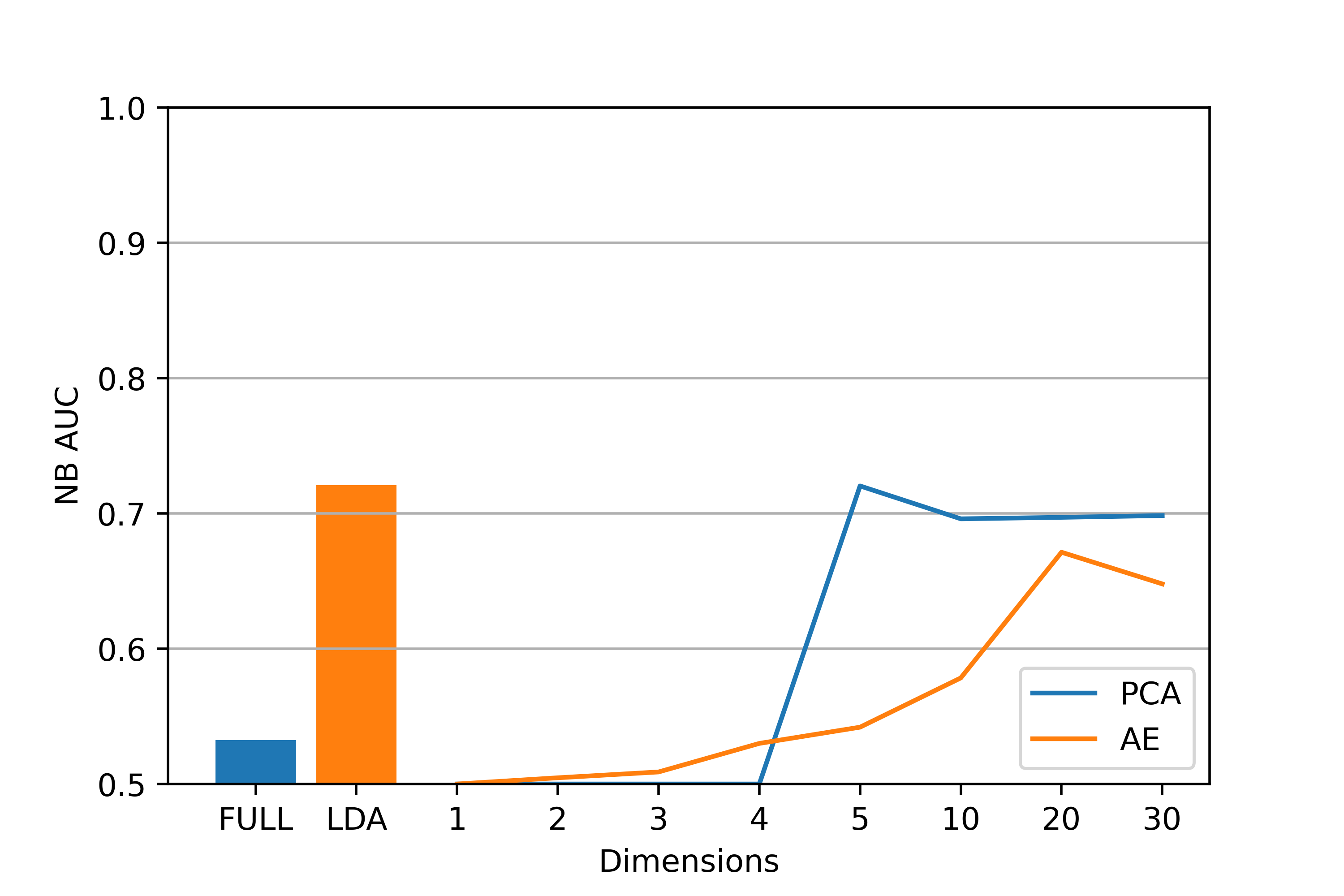}  
  \caption{NB}
  \label{fig:nbton}
\end{subfigure}
\hfill
\begin{subfigure}{.3\textwidth}
  \centering
  \includegraphics[width=5.5cm, height=3cm]{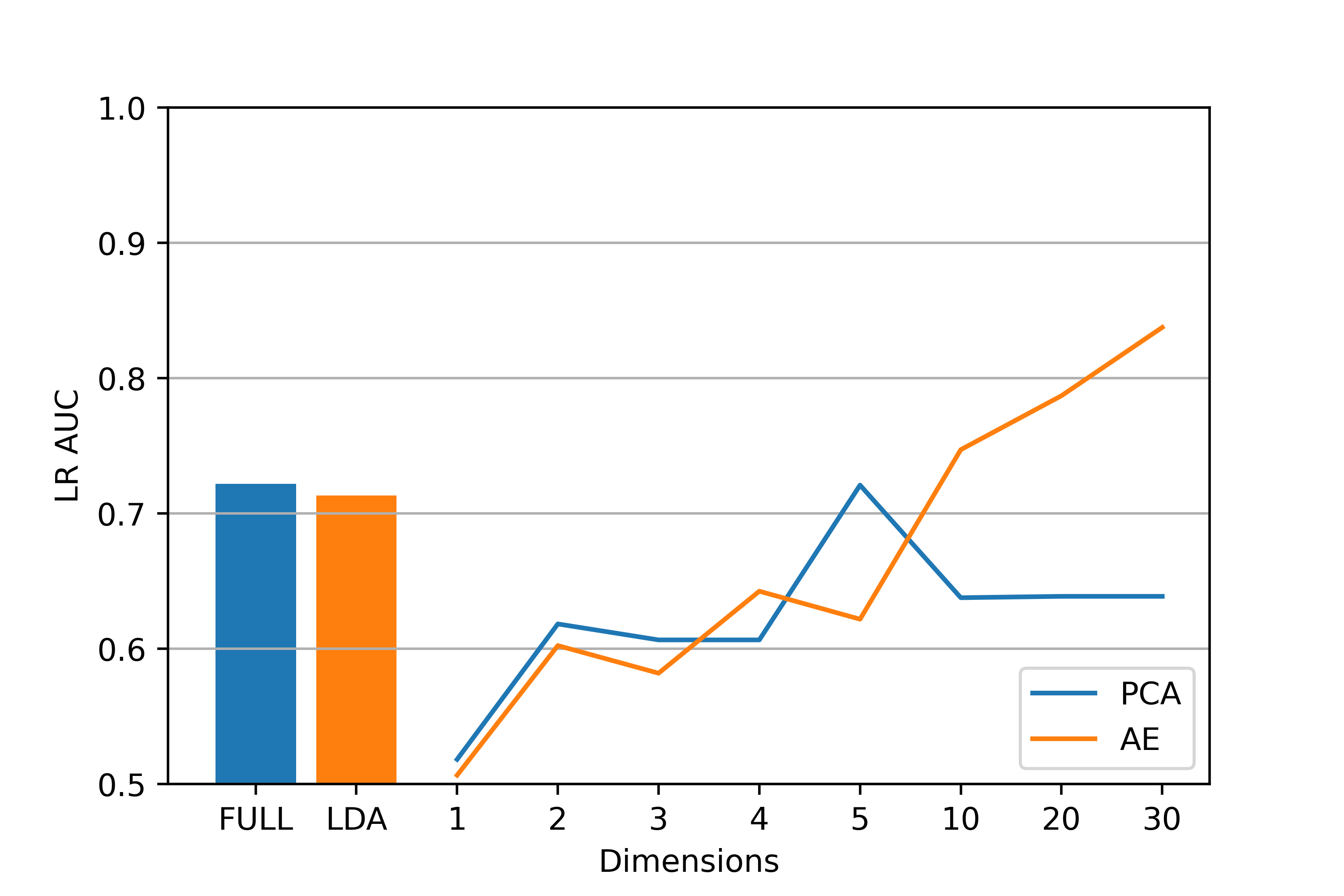}  
  \caption{LR}
  \label{fig:lrton}
\end{subfigure}
\caption{ToN-IoT results}
\label{fig:ton}
\end{figure*}

Using the ToN-IoT dataset, the results achieved by each FE algorithm and ML model are significantly different, as displayed in Fig.\ref{fig:ton}. Overall, DT obtains the best possible results when applied to full dataset, and AE is used. The DFF model achieves its best results on the complete full dataset, performing poorly when using AE but better when using LDA and PCA as it is stable after 4 dimensions. For any dimension less than 10, CNN performs inefficiently with AE and PCA. Like DFF, RNN performs poorly when using AE but well when using PCA as it starts to stabilise with 2 dimensions. DT achieves great results when applied to the full datasets, similar to AE, with dimensions greater than or equal to 2. However, when using LDA or PCA, it will generate defective results. LR and NB do not perform efficiently on this dataset using any of the FE algorithms. LDA improves the performances of RNN and NB but reduces those of DFF, CNN, and DT applied to the full dataset.

\begin{table}[ht]\scriptsize
\centering
\caption{ToN-IoT classification metrics}\label{Metrics2}
\resizebox{\columnwidth}{!}{%
\begin{tabular}{|c|lllllll|}
\hline
\textbf{ML} &
  \multicolumn{1}{c}{\textbf{FE}} &
  \multicolumn{1}{c}{\textbf{DIM}} &
  \multicolumn{1}{c}{\textbf{ACC}} &
  \multicolumn{1}{c}{\textbf{F1}} &
  \multicolumn{1}{c}{\textbf{DR}} &
  \multicolumn{1}{c}{\textbf{FAR}} &
  \multicolumn{1}{c|}{\textbf{AUC}} \\ \hline
\multirow{4}{*}{DFF} & FULL & 37 & 95.45\% & 0.84 & 76.67\% & 1.26\%  & 0.9337 \\ \cline{2-8} 
                     & LDA & 1  & 94.27\% & 0.97 & 95.06\% & 28.32\% & 0.8953 \\ \cline{2-8} 
                     & PCA & 5  & 95.97\% & 0.98 & 96.91\% & 30.94\% & 0.9078 \\ \cline{2-8} 
                     & AE  & 30 & 96.93\% & 0.98 & 98.25\% & 40.86\% & 0.8010  \\ \hline
\multirow{4}{*}{CNN} & FULL & 37 & 95.43\% & 0.98 & 96.29\% & 29.32\% & 0.9100 \\ \cline{2-8} 
                     & LDA & 1  & 97.60\%  & 0.99 & 99.46\% & 55.37\% & 0.8155 \\ \cline{2-8} 
                     & PCA & 10 & 96.44\% & 0.98 & 97.29\% & 27.68\% & 0.9232 \\ \cline{2-8} 
                     & AE  & 20 & 96.78\% & 0.98 & 97.59\% & 26.39\% & 0.9254 \\ \hline
\multirow{4}{*}{RNN} & FULL & 37 & 86.35\% & 0.93 & 87.40\%  & 43.80\%  & 0.7868 \\ \cline{2-8} 
                     & LDA & 1  & 93.03\% & 0.96 & 93.77\% & 28.19\% & 0.8801 \\ \cline{2-8} 
                     & PCA & 5  & 96.13\% & 0.98 & 96.90\% & 26.02\% & 0.9249 \\ \cline{2-8} 
                     & AE  & 4  & 96.02\% & 0.98 & 96.93\% & 30.18\% & 0.9079 \\ \hline
\multirow{4}{*}{LR}  & FULL & 37 & 75.46\% & 0.86 & 75.70\%  & 31.36\% & 0.7217 \\ \cline{2-8} 
                     & LDA & 1  & 97.68\% & 0.99 & 99.59\% & 56.97\% & 0.7131 \\ \cline{2-8} 
                     & PCA & 5  & 75.44\% & 0.86 & 75.68\% & 31.49\% & 0.7209 \\ \cline{2-8} 
                     & AE  & 30 & 95.46\% & 0.98 & 96.31\% & 28.81\% & 0.8375 \\ \hline
\multirow{4}{*}{DT}  & FULL & 37 & 97.29\% & 0.99 & 97.29\% & 2.66\%  & 0.9731 \\ \cline{2-8} 
                     & LDA & 1  & 86.61\% & 0.92 & 87.77\% & 46.53\% & 0.7062 \\ \cline{2-8} 
                     & PCA & 3  & 80.86\% & 0.89 & 81.16\% & 27.92\% & 0.7662 \\ \cline{2-8} 
                     & AE  & 20 & 98.23\% & 0.99 & 98.28\% & 3.21\%  & 0.9753 \\ \hline
\multirow{4}{*}{NB}  & FULL & 37 & 96.78\% & 0.98 & 99.93\% & 93.41\% & 0.5326 \\ \cline{2-8} 
                     & LDA & 1  & 97.77\% & 0.99 & 99.64\% & 55.48\% & 0.7208 \\ \cline{2-8} 
                     & PCA & 5  & 97.94\% & 0.99 & 99.82\% & 55.75\% & 0.7203 \\ \cline{2-8} 
                     & AE  & 20 & 91.47\% & 0.95 & 93.24\% & 58.98\% & 0.6713 \\ \hline
\end{tabular}
}
\end{table}

The full metrics of the best results obtained by each FE method using all ML models on the ToN-IoT dataset are listed in Table \ref{Metrics2}. The FAR values are considerably high because there are more attack samples than benign samples in the dataset. DFF performs best when applied to the full dataset, achieving a low FAR, i.e., 1.26\%, and a low DR of 76.67\%. AE decreases the performance of DFF even after using the maximum number of dimensions provided. FE algorithms, especially PCA, significantly improve the performances of RNN and NB applied to the full dataset. DT obtains the highest scores when applied to the full dataset, and AE extracted dimensions. The best is to use AE with 10 dimensions as the DR of 98.28\% and FAR of 3.21\% are recorded but ineffective when using PCA and LDA. LR and NB achieve the worst performances of the six ML models. LDA proves unreliable compared to PCA and AE for all learning models except RNN and NB.

\begin{table}[ht]\scriptsize
\centering
\caption{ToN-IoT attacks detection}
\label{ToN-IoT Multi}
\begin{tabular}{|l|c|c|c|}
\hline
\textbf{Attack Type} & \textbf{Actual} & \textbf{Predicted} & \textbf{DR} \\ \hline
Backdoor          & 505385          & 505256             & 99.97\%            \\
DDoS              & 6082893         & 6010012            & 98.80\%            \\
DoS               & 1815909         & 1814699            & 99.93\%               \\
Injection         & 452659          & 442137             & 97.68\%             \\
MITM              & 1043            & 773                & 74.11\%               \\
Password          & 1365958         & 1359372            & 99.52\%                   \\
Ransomware        & 32214           & 10781              & 33.47\%            \\
Scanning          & 7140158         & 6974943            & 97.69\%                 \\
XSS               & 2108944         & 2084863            & 98.86\%               \\ \hline
\end{tabular}
\end{table}

DFF, RNN, LR and NB obtain their best results for PCA using 5 dimensions, making it the best number of dimensions, while AE requires a higher number of 20. Table \ref{ToN-IoT Multi} displays the types of attacks in this dataset and their actual number of samples compared with the number of classified ones. The best-performing combination of FE and ML methods has been used for prediction, and DT is applied to an AE of 20 dimensions having a 98.28\% DR. This table shows that each attack type is almost fully detected except for MITM and ransomware because there are few samples of each of the models to train on. Scanning and injection attacks have 97.69\% and 97.68\% DRs, respectively, despite their sufficient samples, indicating that their patterns are more complex.


\subsection{CSE-CIC-IDS2018}
\begin{figure*}[ht]
\begin{subfigure}{.3\textwidth}
  \centering
  \includegraphics[width=5.5cm, height=3cm]{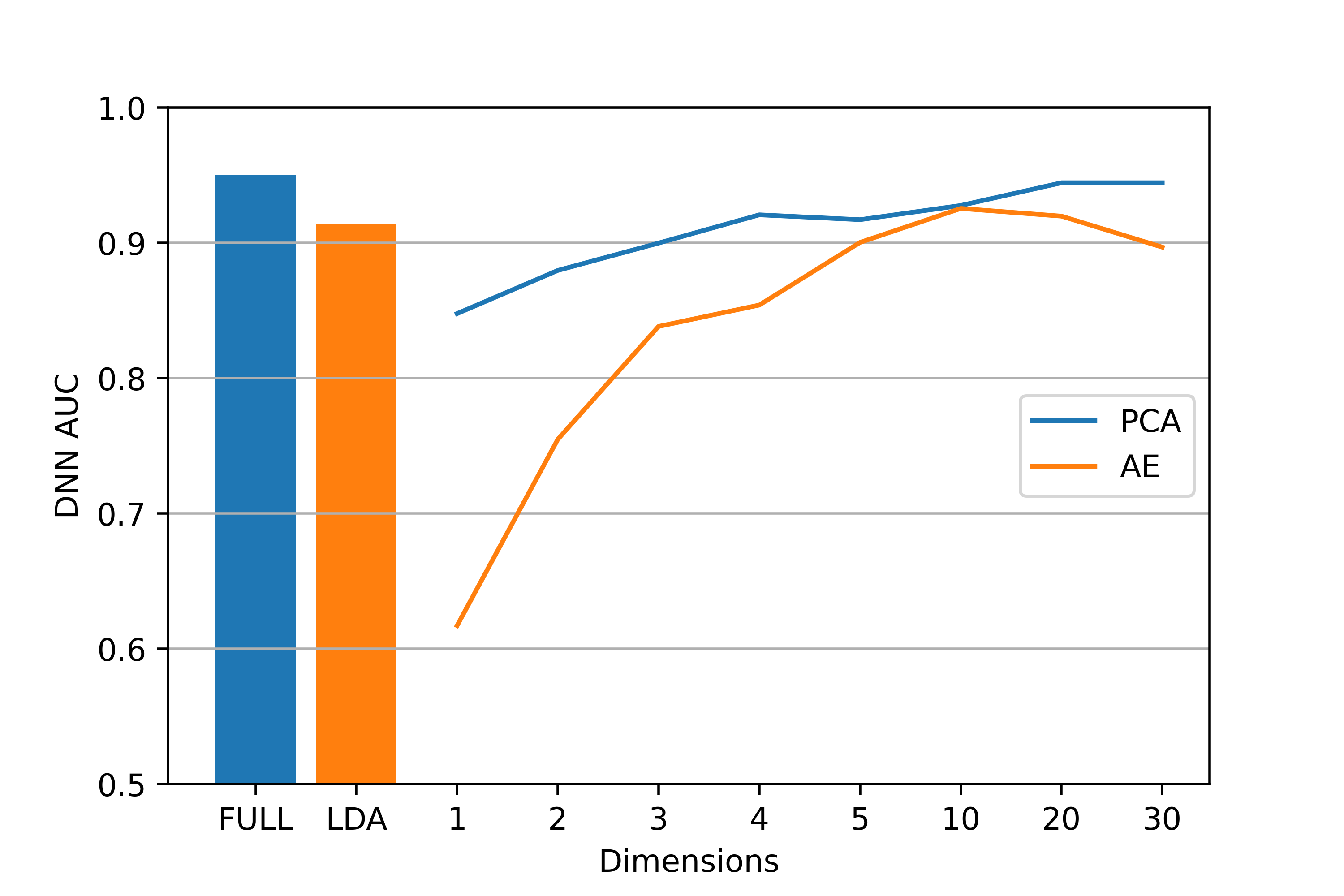}  
  \caption{DFF}
  \label{fig:dnncse}
\end{subfigure}
\hfill
\begin{subfigure}{.3\textwidth}
  \centering
  \includegraphics[width=5.5cm, height=3cm]{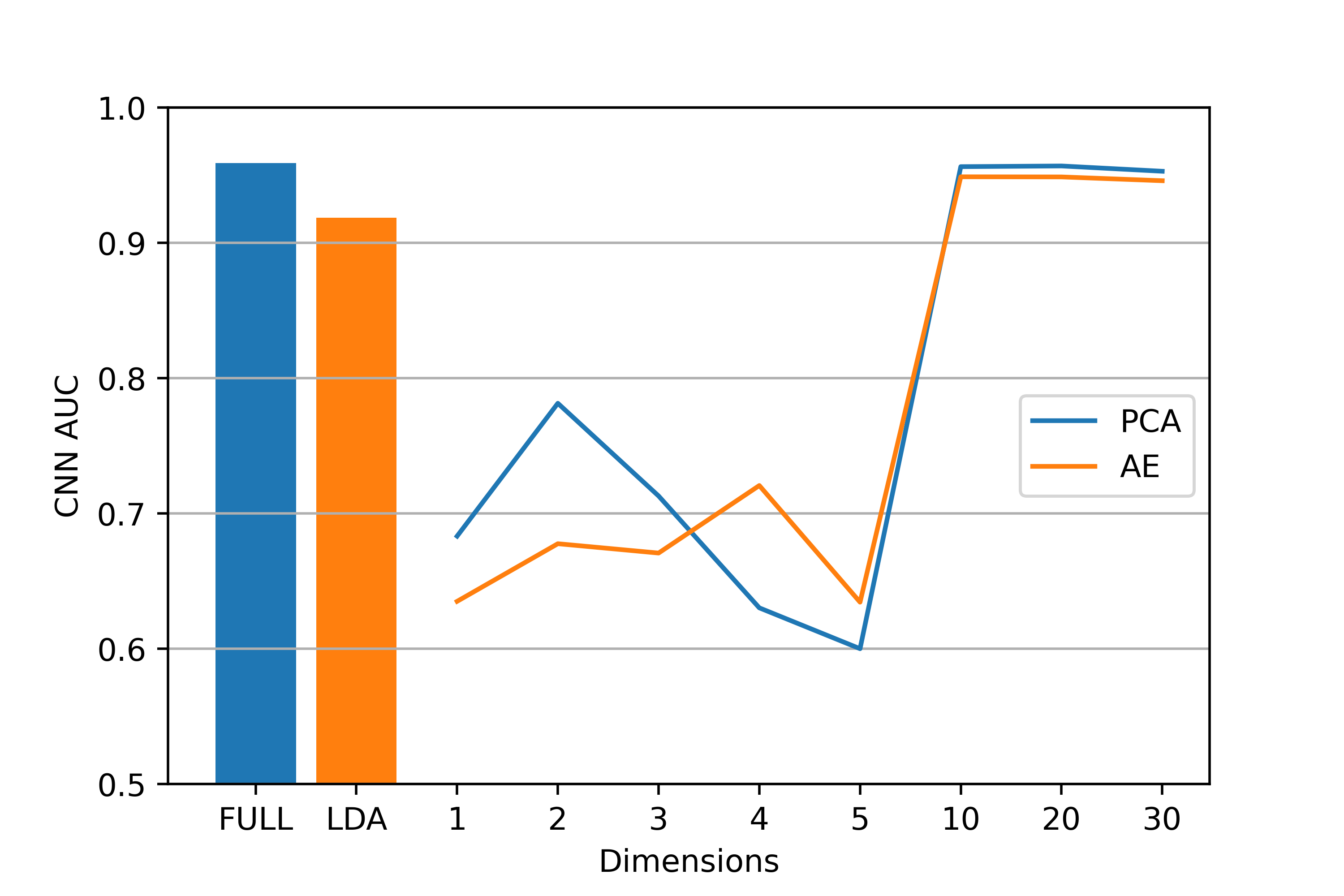}  
  \caption{CNN}
  \label{fig:cnncse}
\end{subfigure}
\hfill
\begin{subfigure}{.3\textwidth}
  \centering
  \includegraphics[width=5.5cm, height=3cm]{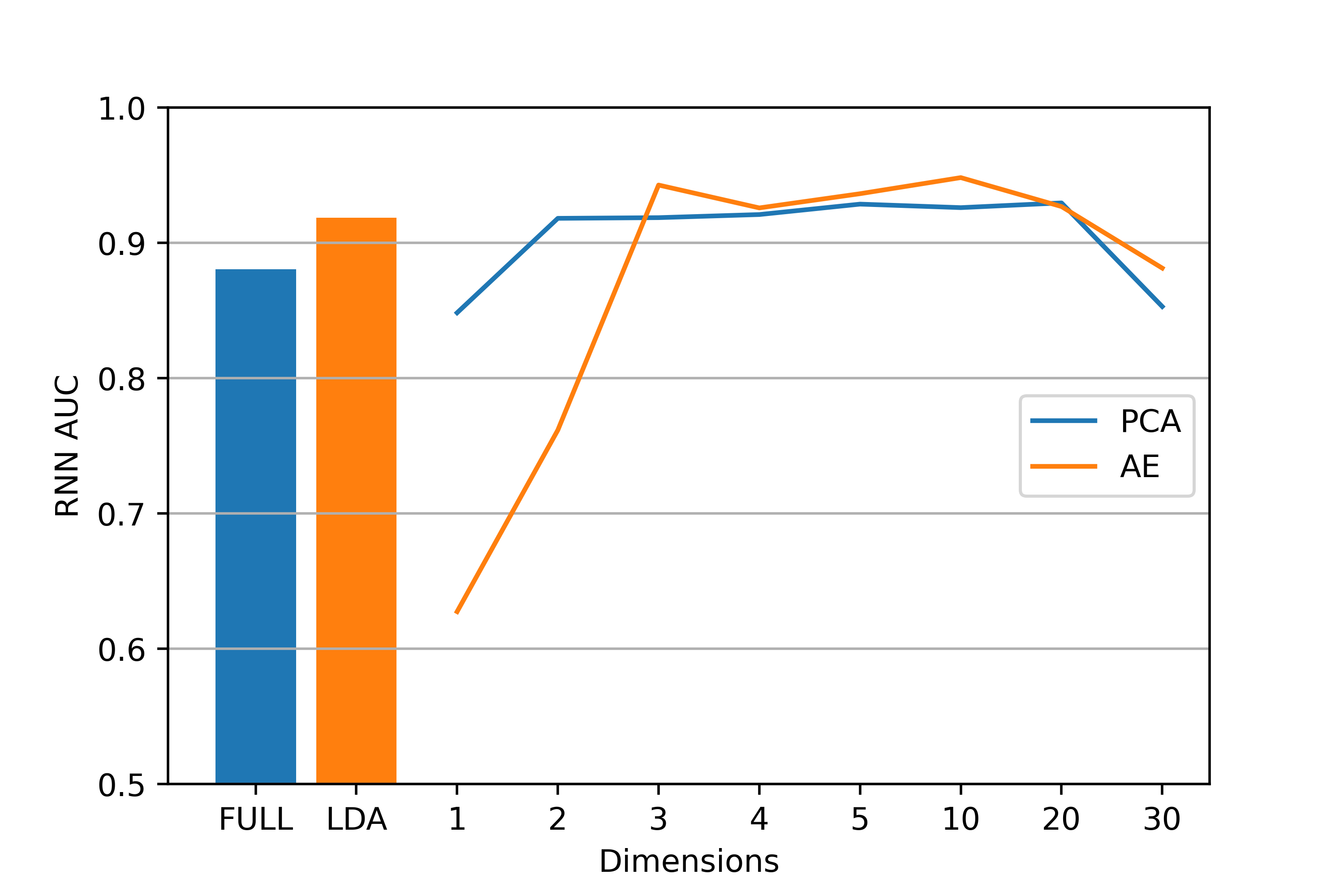}  
  \caption{RNN}
  \label{fig:rnncse}
\end{subfigure}
\hfill
\begin{subfigure}{.3\textwidth}
  \centering
  \includegraphics[width=5.5cm, height=3cm]{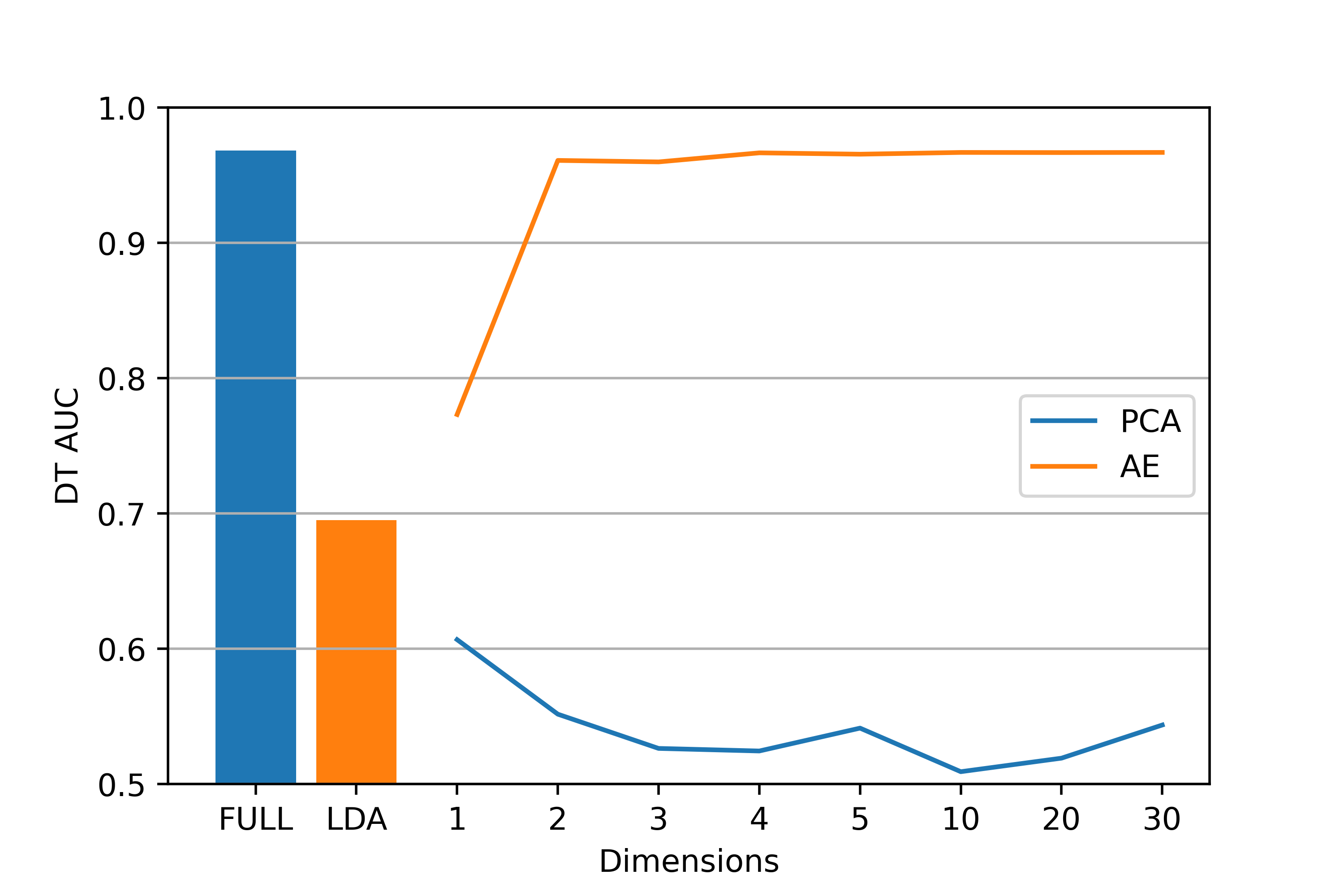}  
  \caption{DT}
  \label{fig:dtcse}
\end{subfigure}
\hfill
\begin{subfigure}{.3\textwidth}
  \centering
  \includegraphics[width=5.5cm, height=3cm]{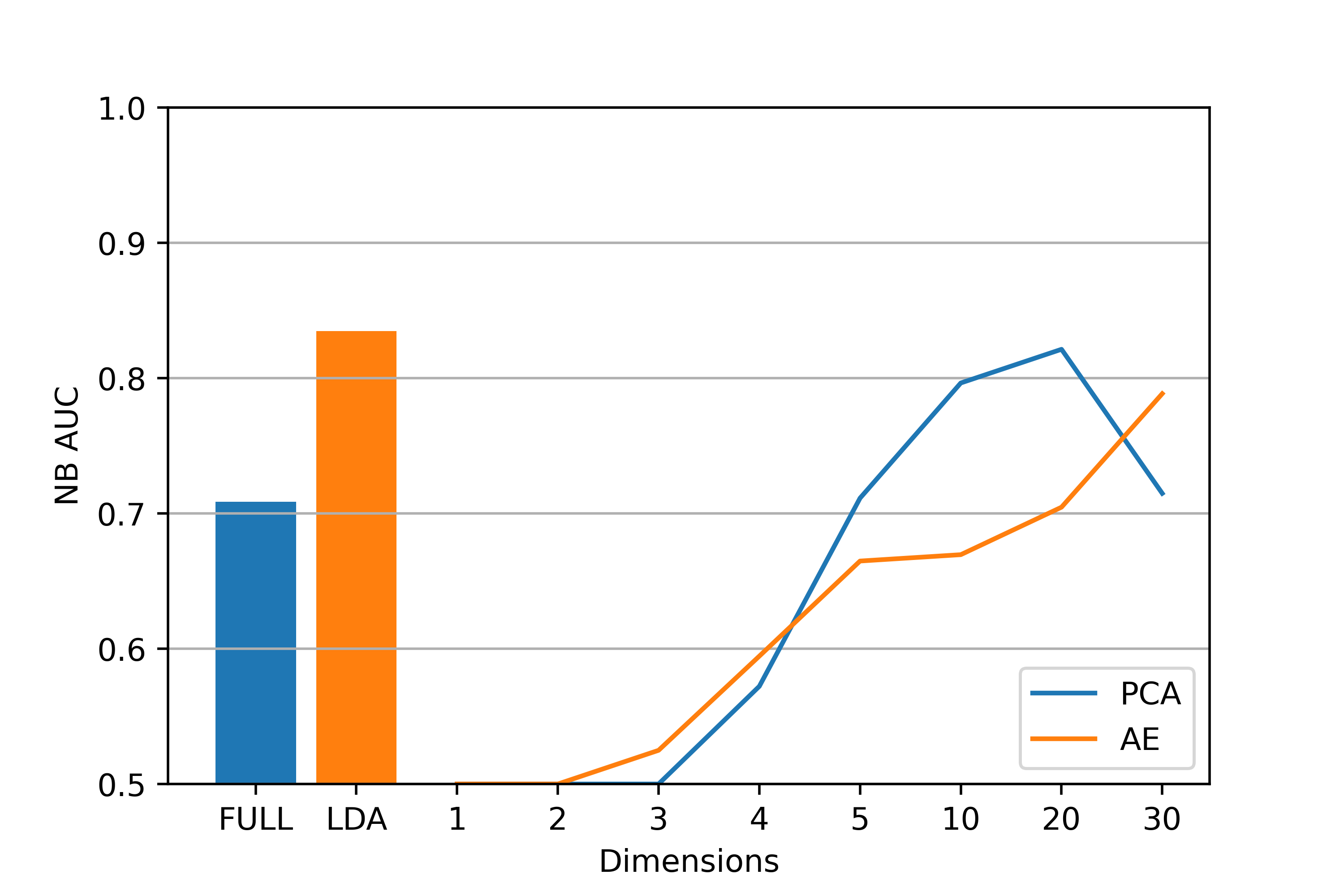}  
  \caption{NB}
  \label{fig:nbcse}
\end{subfigure}
\hfill
\begin{subfigure}{.3\textwidth}
  \centering
  \includegraphics[width=5.5cm, height=3cm]{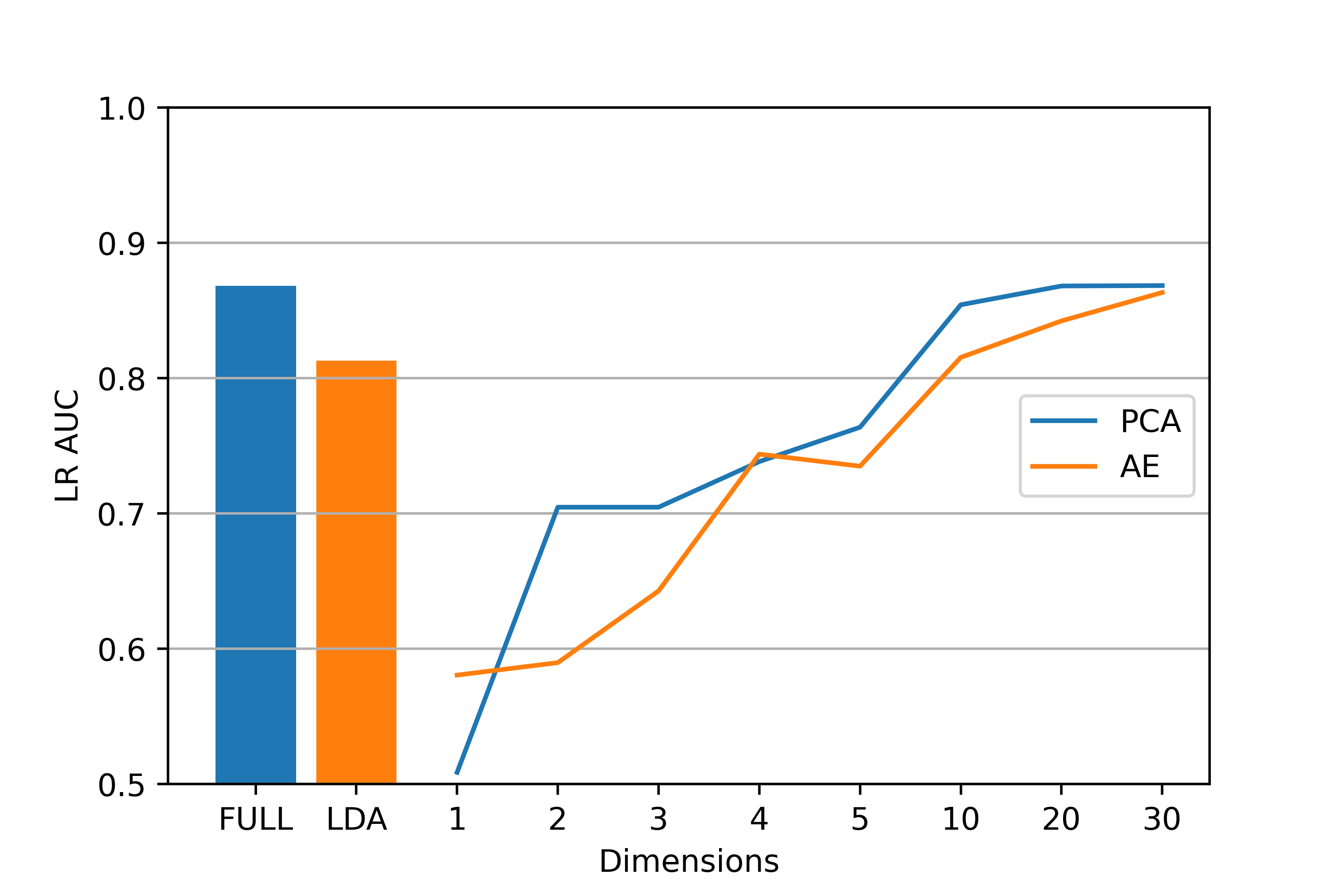}  
  \caption{LR}
  \label{fig:lrcse}
\end{subfigure}
\caption{CSE-CIC-IDS2018 results}
\label{fig:cse-cic-ids}
\end{figure*}

As illustrated in Fig.\ref{fig:cse-cic-ids}, the DL models perform equally well in terms of their best AUC scores. DFF is applied to the full dataset, and good detection performance is achieved when PCA is used. The effects of the AE's and PCA's changing dimensions are very similar for CNN as it also has difficulty in classification using a lower number of dimensions. RNN performs equally using all FE algorithms, with AE slightly better than others. DT performs well with AE and when applied to the full dataset but performs very poorly with LDA and PCA. Using AE requires only 3 dimensions to stabilise and reach its maximum AUC. NB obtains its best results using LDA and PCA, peaking at dimension 20, and LR performs equally using the three FE algorithms. Moreover, AE and PCA have similar impacts on all ML models except DT, for which AE significantly outperforms PCA. LR and NB perform poorly throughout the experiments.

\begin{table}[ht]\scriptsize
\centering
\caption{CSE-CIC-IDS2018 classification metrics}\label{Metrics3}
\resizebox{\columnwidth}{!}{%
\begin{tabular}{|c|lllllll|}
\hline
\textbf{ML} &
  \multicolumn{1}{c}{\textbf{FE}} &
  \multicolumn{1}{c}{\textbf{DIM}} &
  \multicolumn{1}{c}{\textbf{ACC}} &
  \multicolumn{1}{c}{\textbf{F1}} &
  \multicolumn{1}{c}{\textbf{DR}} &
  \multicolumn{1}{c}{\textbf{FAR}} &
  \multicolumn{1}{c|}{\textbf{AUC}} \\ \hline
\multirow{4}{*}{DFF} & FULL & 76 & 96.11\% & 0.86 & 81.83\% & 1.39\%  & 0.9504 \\ \cline{2-8} 
                     & LDA & 1  & 91.57\% & 0.72 & 71.46\% & 4.92\%  & 0.9141 \\ \cline{2-8} 
                     & PCA & 20 & 95.49\% & 0.85 & 83.41\% & 2.39\%  & 0.9444 \\ \cline{2-8} 
                     & AE  & 10 & 89.28\% & 0.60  & 58.20\%  & 5.29\%  & 0.9254 \\ \hline
\multirow{4}{*}{CNN} & FULL & 76 & 96.23\% & 0.87 & 85.87\% & 1.95\%  & 0.9590 \\ \cline{2-8} 
                     & LDA & 1  & 90.59\% & 0.71 & 73.58\% & 6.44\%  & 0.9186 \\ \cline{2-8} 
                     & PCA & 20 & 95.16\% & 0.74 & 85.49\% & 3.15\%  & 0.9563 \\ \cline{2-8} 
                     & AE  & 10 & 96.72\% & 0.89 & 85.73\% & 1.36\%  & 0.9487 \\ \hline
\multirow{4}{*}{RNN} & FULL & 76 & 86.93\% & 0.64 & 73.20\%  & 10.67\% & 0.8806 \\ \cline{2-8} 
                     & LDA & 1  & 91.85\% & 0.73 & 73.65\% & 4.97\%  & 0.9187 \\ \cline{2-8} 
                     & PCA & 20 & 85.00\%    & 0.66 & 93.10\%  & 16.42\% & 0.9295 \\ \cline{2-8} 
                     & AE  & 10 & 94.42\% & 0.82 & 82.52\% & 3.50\%   & 0.9482 \\ \hline
\multirow{4}{*}{LR}  & FULL & 76 & 81.74\% & 0.60  & 94.85\% & 21.22\% & 0.8681 \\ \cline{2-8} 
                     & LDA & 1  & 82.39\% & 0.57 & 79.75\% & 17.14\% & 0.8130 \\ \cline{2-8} 
                     & PCA & 20 & 82.22\% & 0.61 & 93.43\% & 19.74\% & 0.8684 \\ \cline{2-8} 
                     & AE  & 30 & 82.08\% & 0.61 & 92.37\% & 19.72\% & 0.8633 \\ \hline
\multirow{4}{*}{DT}  & FULL & 76 & 98.15\% & 0.94 & 94.94\% & 1.29\%  & 0.9683 \\ \cline{2-8} 
                     & LDA & 1  & 86.96\% & 0.50  & 44.62\% & 5.64\%  & 0.6949 \\ \cline{2-8} 
                     & PCA & 1  & 86.11\% & 0.33 & 24.45\% & 3.11\%  & 0.6067 \\ \cline{2-8} 
                     & AE  & 10 & 98.02\% & 0.93 & 94.76\% & 1.41\%  & 0.9668 \\ \hline
\multirow{4}{*}{NB}  & FULL & 76 & 52.83\% & 0.38 & 96.51\% & 54.80\%  & 0.7086 \\ \cline{2-8} 
                     & LDA & 1  & 93.59\% & 0.76 & 69.09\% & 2.12\%  & 0.8348 \\ \cline{2-8} 
                     & PCA & 20 & 72.97\% & 0.51 & 95.17\% & 30.91\% & 0.8213 \\ \cline{2-8} 
                     & AE  & 30 & 69.17\% & 0.47 & 92.63\% & 34.93\% & 0.7885 \\ \hline
\end{tabular}
}
\end{table}

Table \ref{Metrics3} displays the best score obtained by the FE algorithms for each ML model applied to the CSE-CIC-IDS2018 dataset. DFF and CNN achieve their best performances when applied to the full dataset, while the FE algorithms improve the classification capability of RNNs. LDA performs worse than AE and PCA for all models except NB. However, LR and NB are ineffective in detecting attacks present in this dataset. The optimal numbers of PCA and AE dimensions are 20 and 10, respectively, due to their requirement in most ML classifiers. In Table \ref{CSE-CIC-IDS Multi}, attack types in the dataset and their actual numbers compared with their correct predictions are presented. The best-performing combination of the model and FE algorithm has been used for prediction; that is, the DT classifier is applied to 10 extracted dimensions using AE. This table shows that each attack type is almost fully detected, except Brute Force -Web, Brute Force -XSS, and SQL injection, due to their low number of sample counts in the dataset, which matches the findings in \cite{li_chen_zhang_wu_2020}. However, infiltration attacks are more difficult to detect despite their majority in the dataset. This could be due to the similarity of its statistical distribution with another class type, leading to confusion of the detection model. Further analysis is required, such as t-tests, to measure the difference between the distributions of each class.

\begin{table}[ht]\scriptsize
\centering
\caption{CSE-CIC-IDS2018 attacks detection}
\label{CSE-CIC-IDS Multi}
\begin{tabular}{|l|c|c|c|c|}
\hline
\textbf{Attack Type}         & \textbf{Actual} & \textbf{Predicted} & \textbf{DR}  \\ \hline
Bot                       & 282310          & 282064             & 99.89\%             \\
Brute Force -Web          & 611             & 425                & 60.23\%                 \\
Brute Force -XSS          & 230             & 198                & 79.36\%                \\
DDOS attack -HOIC         & 668461          & 668461             & 100\%                  \\
DDOS attack -LOIC-UDP     & 1730            & 1730               & 99.71\%               \\
DDoS attacks -LOIC-HTTP   & 576191          & 576157             & 99.99\%           \\
DoS attacks -GoldenEye    & 41455           & 41449              & 98.28\%                \\
DoS attacks -Hulk         & 434873          & 434867             & 99.99\%                 \\
DoS attacks -SlowHTTPTest & 19462           & 19462              & 100\%                   \\
DoS attacks -Slowloris    & 10285           & 10190              & 98.86\%                 \\
FTP-BruteForce            & 39352           & 39352              & 100\%                  \\
Infilteration             & 161792          & 43315              & 24.77\%               \\
SQL Injection             & 87              & 73                 & 41.245\%                  \\
SSH-Bruteforce            & 117322          & 117321             & 99.28\%           \\ \hline
\end{tabular}
\end{table}



\subsection{Discussion}

According to the evaluation results, it has been observed that a relatively small number of feature dimensions can achieve classification performance close to the maximum. In addition, the marginal income of more dimensions is very small. The outputs of LDA and PCA are analysed using their respective variance to understand and explain this behaviour. The variance is the distribution of the squared deviations of the output from its respective mean. The variance of each dimension extracted from all the datasets using PCA and LDA is discussed. Measuring the variance of the dimensions being fed into the ML classifiers is necessary for this field. It will aid in understanding how FE techniques perform on NIDS datasets.

\begin{figure}[ht]
  \centering
  \includegraphics[width=7cm, height=3.5cm]{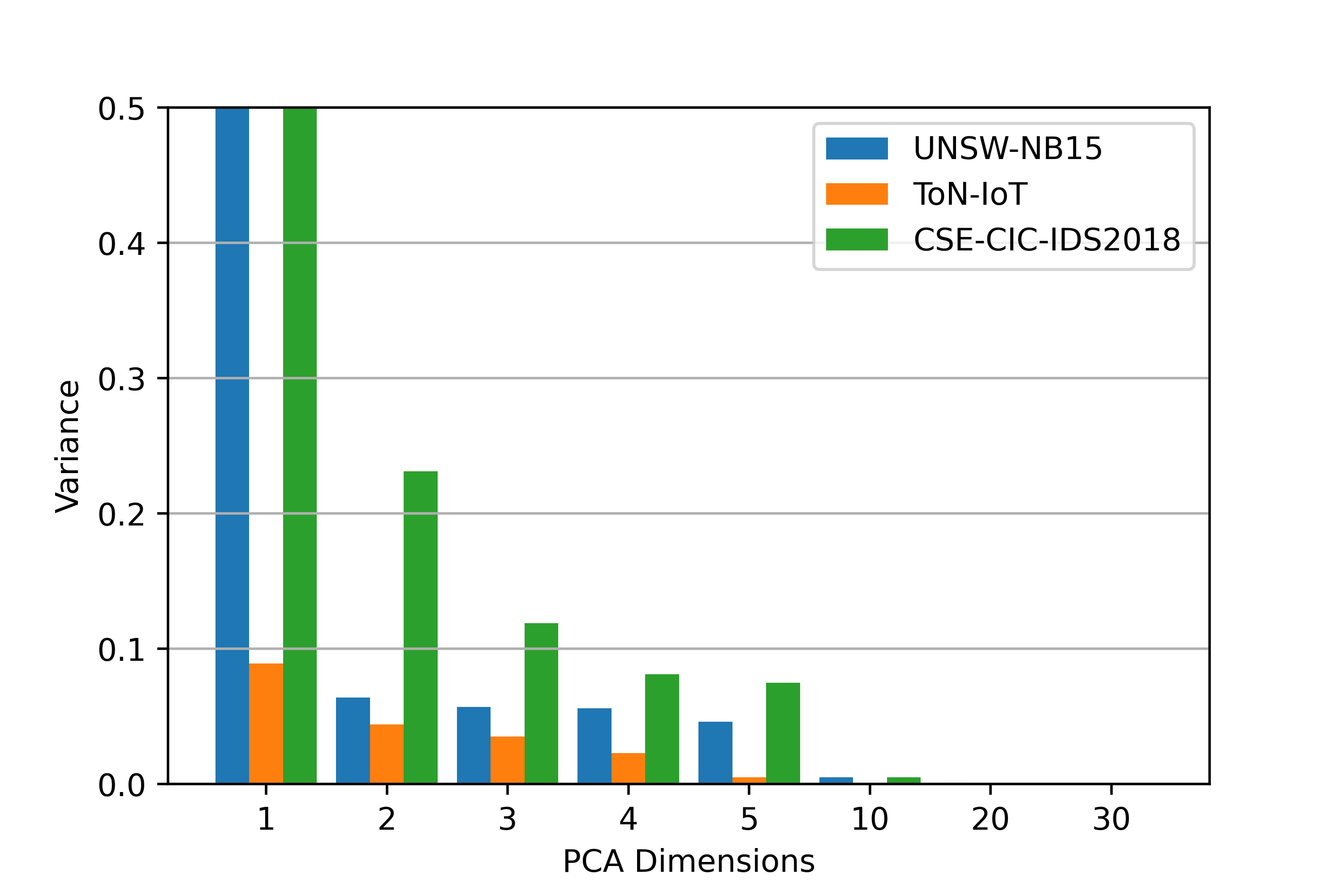}  
  \caption{Variance of the extracted PCA dimensions}
  \label{vpca2}
\end{figure}
  
\begin{figure}[ht]
  \centering
  \includegraphics[width=4cm, height=3.5cm]{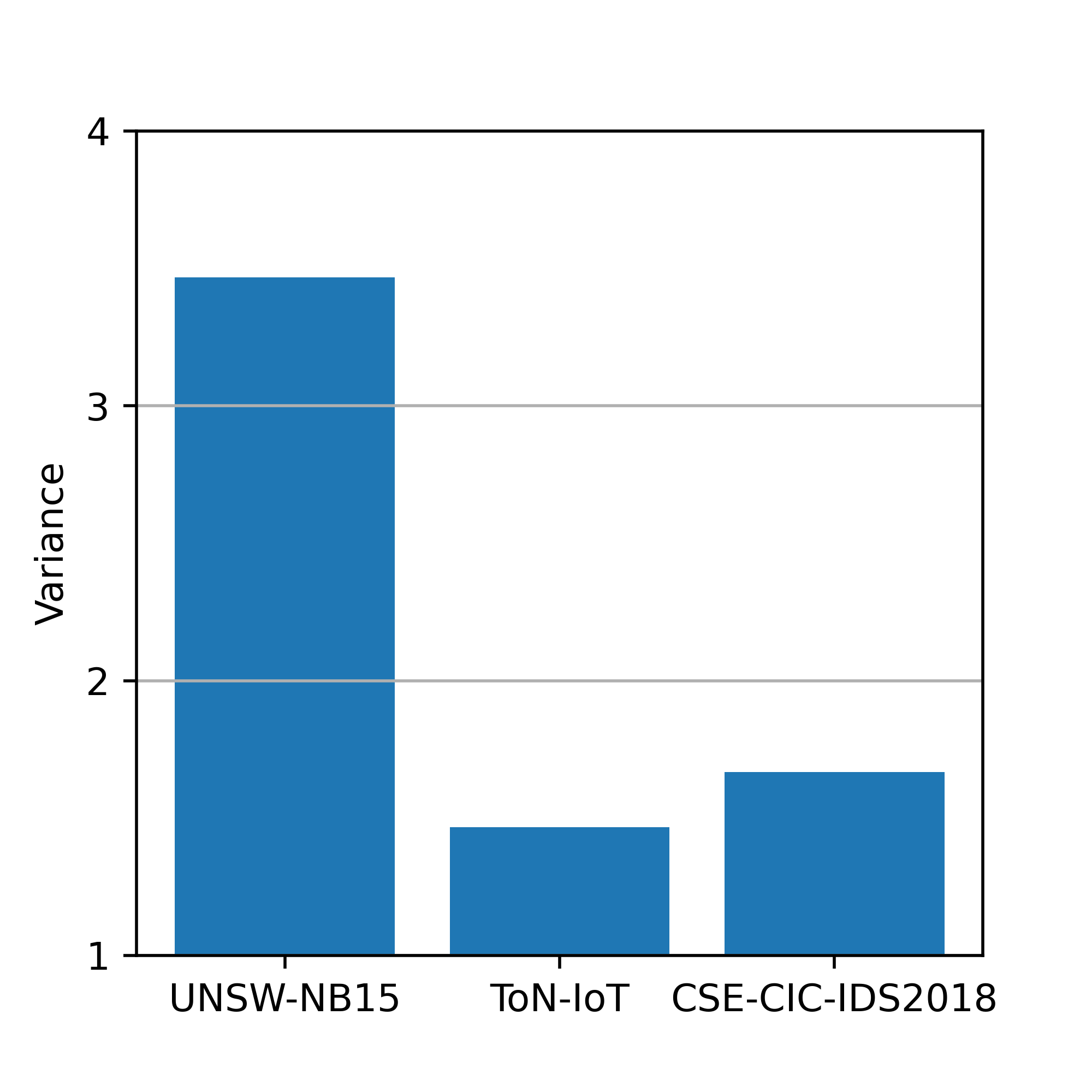}  
  \caption{Variance of the extracted LDA dimension}
  \label{vlda}
\end{figure}

Fig.\ref{vpca2} shows the variance of each dimension extracted in PCA for the three datasets. As observed, the first 10 feature dimensions account for the bulk of the variance, with a minor contribution of additional dimensions. This is consistent with and explains the results in Figs. \ref{fig:fig}, \ref{fig:ton}, and \ref{fig:cse-cic-ids}, where a higher number of features beyond 10 does not provide any further increase in classification accuracy. Fig.\ref{vlda} displays the variance of the single LDA feature for each of the three considered datasets. The extracted LDA feature of the UNSW-NB15 dataset has a significantly higher variance compared to the other two datasets. This might indicate that one or a very small number of features in the UNSW-NB15 dataset strongly correlate to the labels. This is consistent with the results observed in Figs. \ref{fig:fig}, \ref{fig:ton} and \ref{fig:cse-cic-ids}, where the LDA for the UNSW-NB15 dataset achieves a significantly higher classification accuracy than the other two datasets. The classification accuracy of LDA for UNSW-NB15 is close to that achieved with the full dataset, i.e., with the complete set of features.

\begin{figure*}[ht!]
\begin{subfigure}{.3\textwidth}
  \centering
  \includegraphics[width=5.5cm, height=3cm]{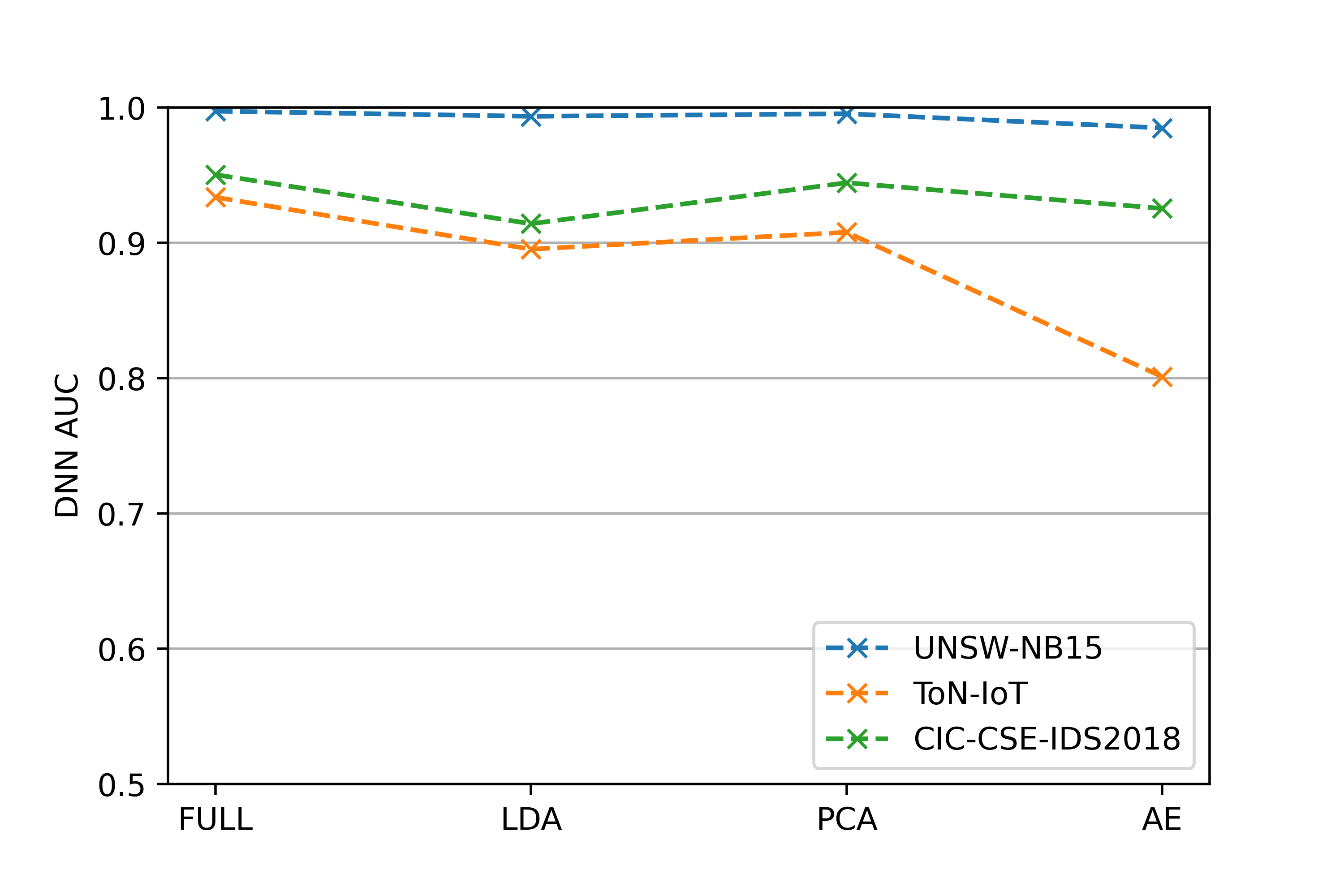}  
  \caption{DFF}
  \label{fig:dnno}
\end{subfigure}
\hfill
\begin{subfigure}{.3\textwidth}
  \centering
  \includegraphics[width=5.5cm, height=3cm]{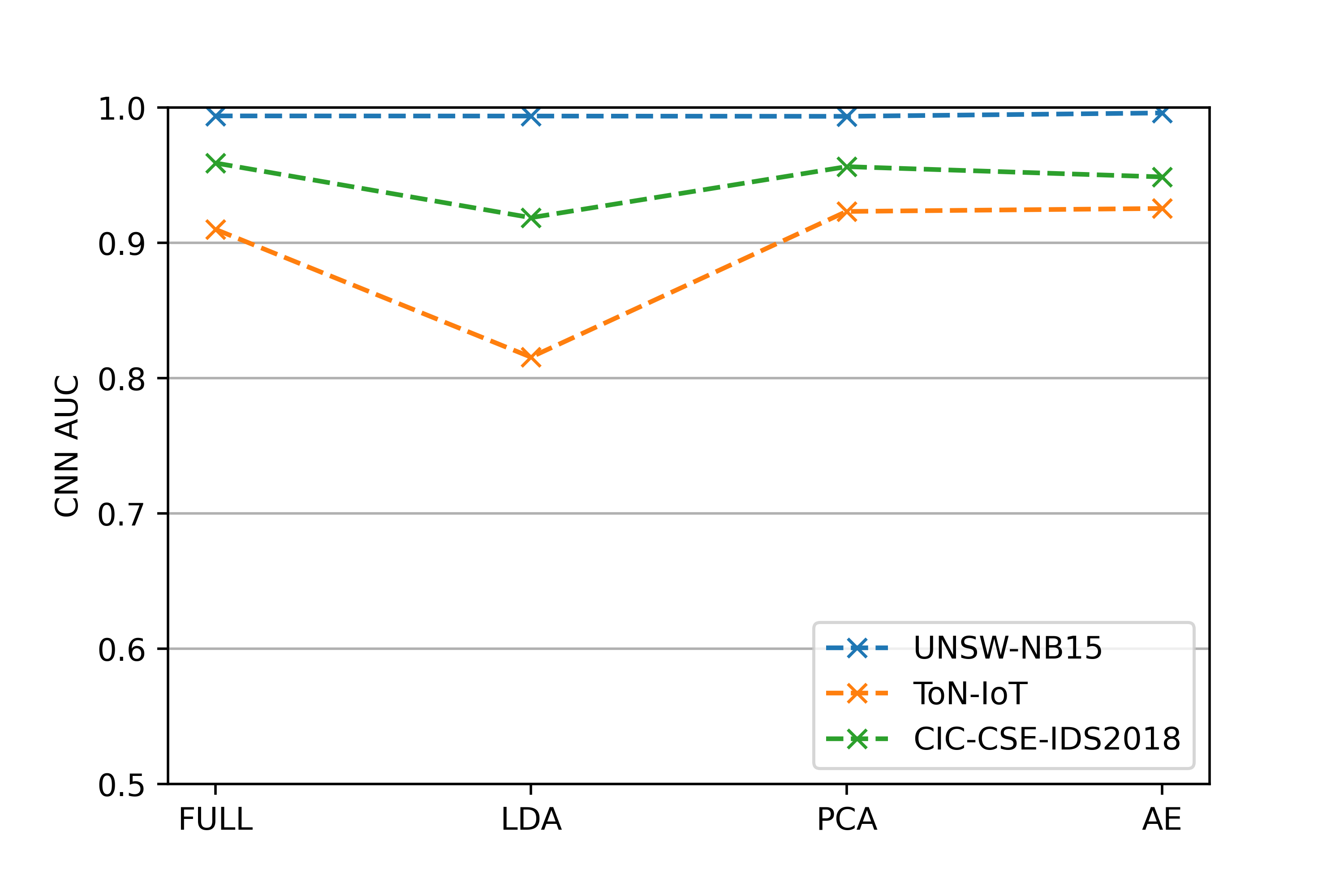}  
  \caption{CNN}
  \label{fig:cnno}
\end{subfigure}
\hfill
\begin{subfigure}{.3\textwidth}
  \centering
  \includegraphics[width=5.5cm, height=3cm]{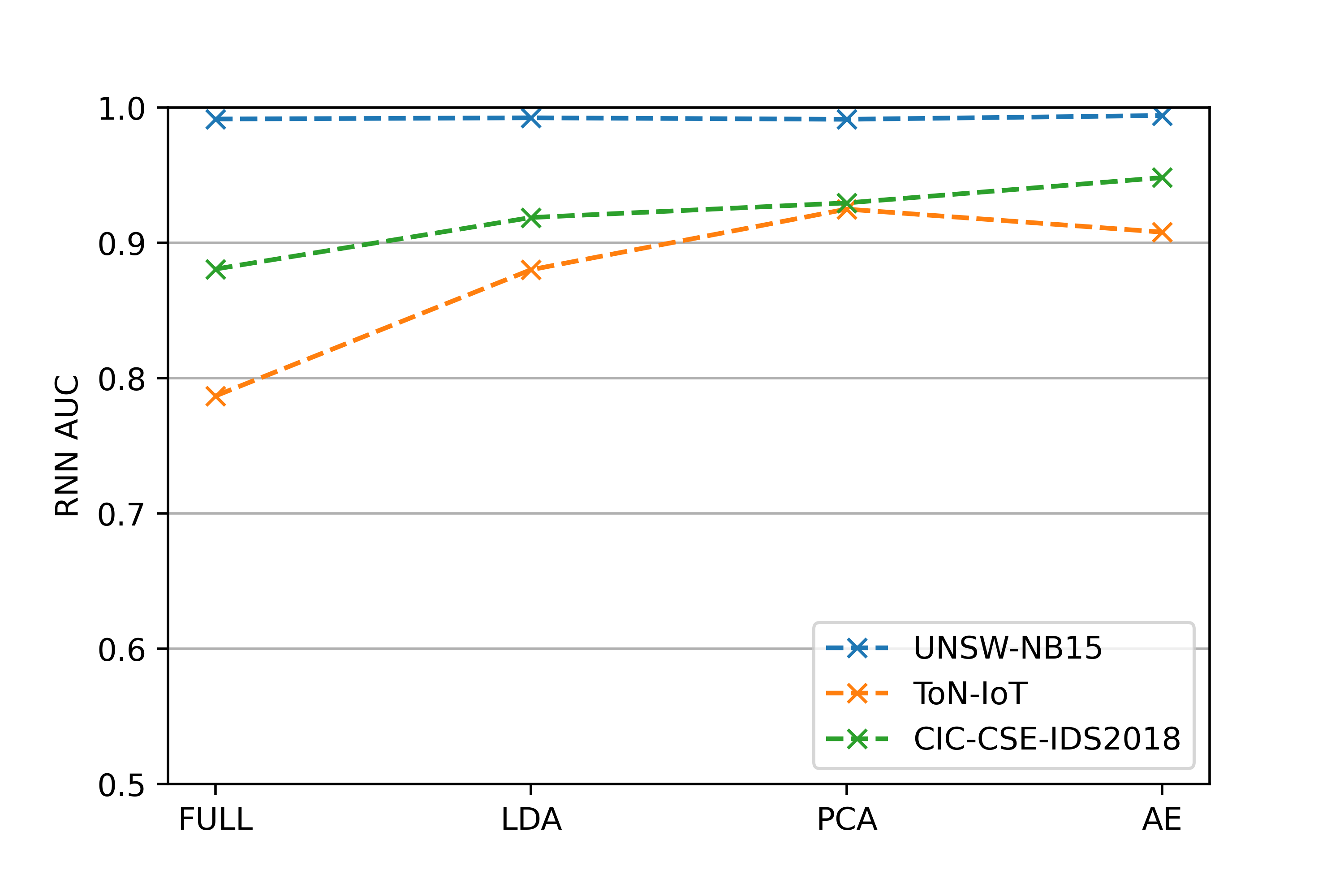}  
  \caption{RNN}
  \label{fig:rnno}
\end{subfigure}
\hfill
\begin{subfigure}{.3\textwidth}
  \centering
  \includegraphics[width=5.5cm, height=3cm]{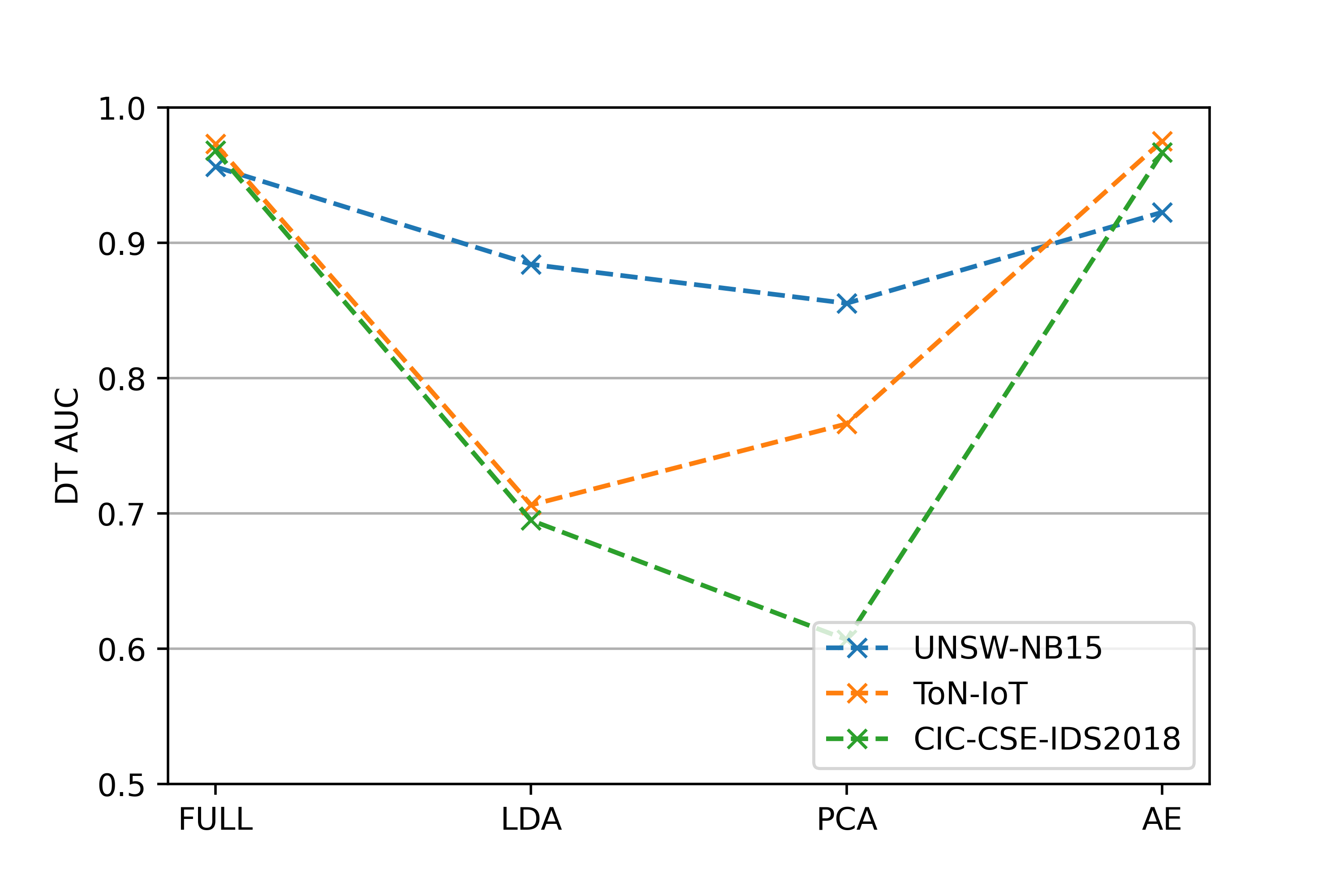}  
  \caption{DT}
  \label{fig:dto}
\end{subfigure}
\hfill
\begin{subfigure}{.3\textwidth}
  \centering
  \includegraphics[width=5.5cm, height=3cm]{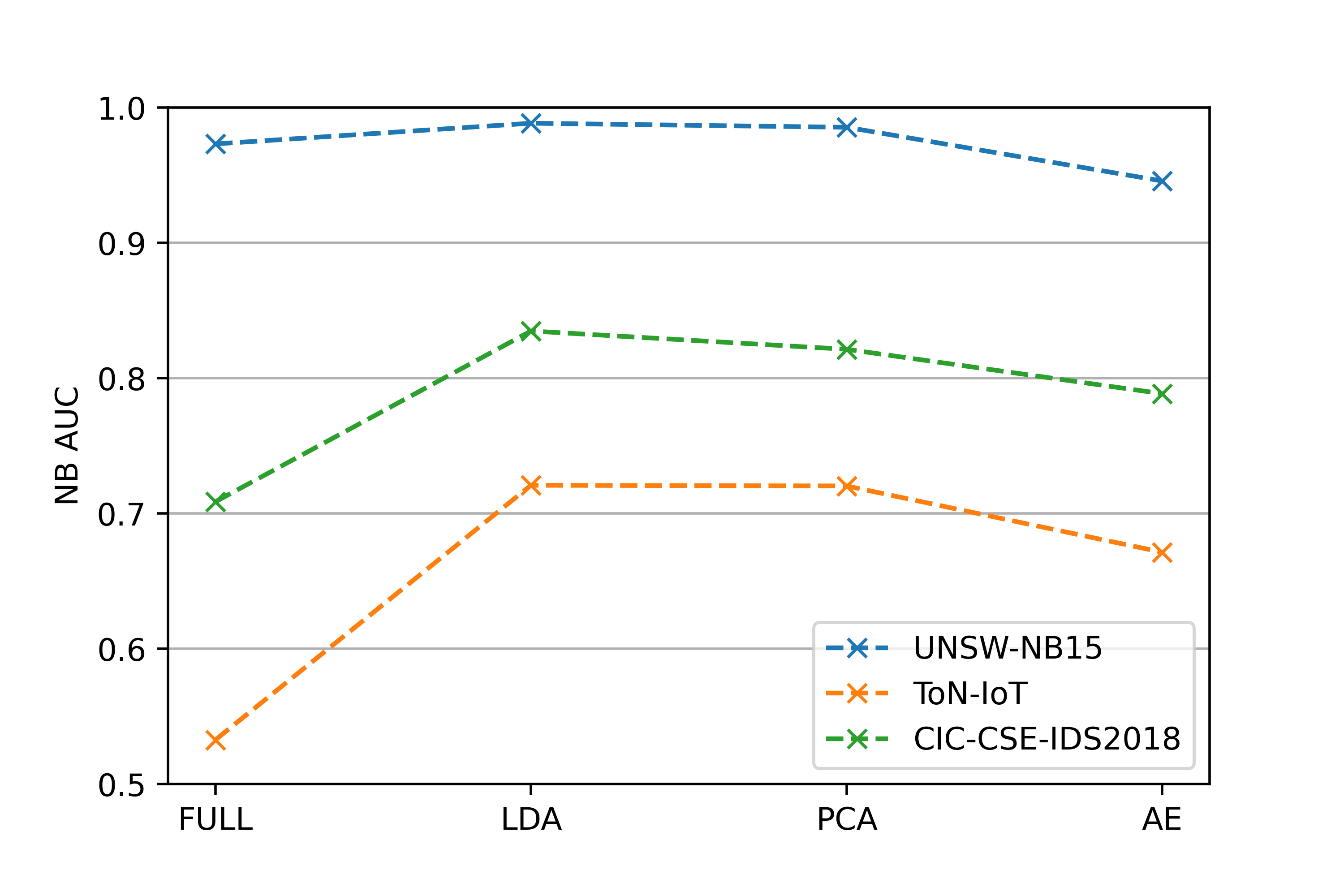}  
  \caption{NB}
  \label{fig:nbo}
\end{subfigure}
\hfill
\begin{subfigure}{.3\textwidth}
  \centering
  \includegraphics[width=5.5cm, height=3cm]{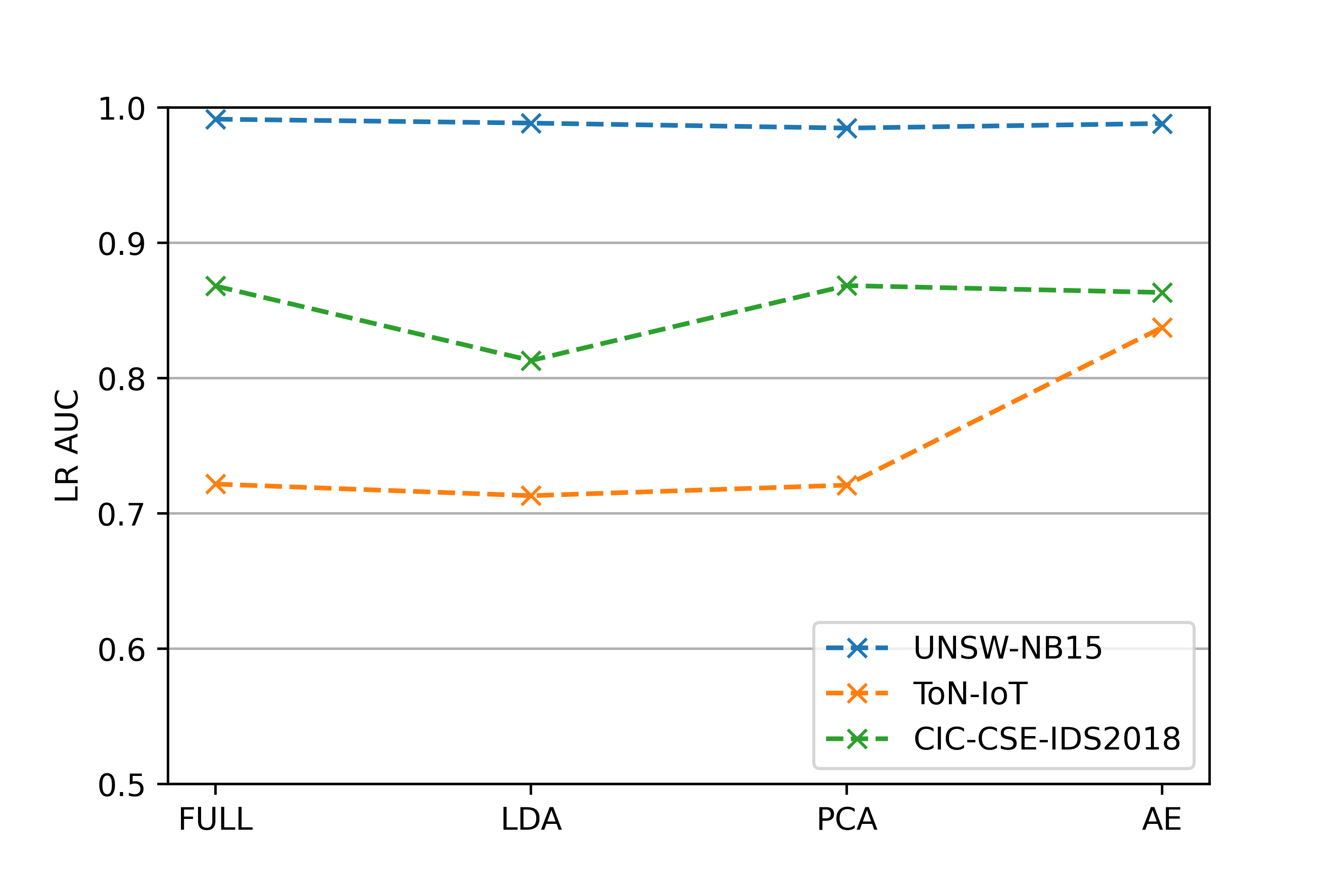}  
  \caption{LR}
  \label{fig:lro}
\end{subfigure}
\caption{Performance of ML classifiers across three NIDS datasets}
\label{fig:summary}
\end{figure*}

The results for the datasets have been grouped based on the ML models, as shown in Fig.\ref{fig:summary}. The best dimensions of PCA and AE are selected for a fair comparison. It is clear that patterns form the effects of applying FE algorithms. In Fig.\ref{fig:summary}(a), DFF is the best when applied to the full dataset due to the ability of a dense network to assign weights to relevant features, while AE lowers the detection accuracy of the DFF model. Figure Fig.\ref{fig:summary}(b) shows that applying CNN to the full dataset or using PCA or AE does not significantly alter its performance, but using LDA, the outcome deteriorates. In Fig.\ref{fig:summary}(c), the necessity of applying an FE algorithm before using RNN is obvious, with the best being AE, followed by PCA, and lastly, LDA. Fig.\ref{fig:summary}(d) proves the unreliability of using LDA or PCA for a DT model, whereas this model works efficiently when applied to the full dataset or using AE. In Fig.\ref{fig:summary}(e), applying a linear FE algorithm, namely, LDA or PCA, improves the performance of the NB model. LDA achieves the best results, while the NB has the worst results among the six ML models without an FE algorithm. Fig.\ref{fig:summary}(f) shows that applying LR to the full dataset or using FE methods leads to the same results where AE improves the model's performance on the ToN-IoT dataset while LDA decreases it on the CSE-CIC-IDS2018. Overall, there is a clear pattern of the effects of the FE methods and classification capabilities of ML models for the three datasets. Models such as RNN and NB benefit from applying FE algorithms, whereas DFF does not. LDA's general performance is negative for the ToN-IoT and CSE-CIC-IDS2018 datasets when using all ML models except NB. This is explained by the low variance scores achieved by the two datasets compared to the UNSW-NB15 dataset. However, LR and NB do not perform well for detecting attacks in the three datasets, with the best scores attained by a different set of techniques.

The experimental evaluation of 18 different combinations of FE and ML techniques has assisted in finding the optimal combination for each dataset used. On the UNSW-NB15 dataset, the CNN classifier obtains the best score when applied to the AE dimensions. On the ToN-IoT and CSE-CIC-IDS2018 ones, DT outperforms the other models and achieves the best scores using the AE technique. However, no single method works best across the utilised NIDS datasets. This is caused by the vast difference in the feature sets that make up the utilised datasets. Therefore, creating a universal set of features for future NIDS datasets is essential. The universal set needs to be easily generated from live network traffic headers as they do not require deep packet inspection, which is challenging in encrypted traffic. The features should also not be biased towards providing information on limited protocols or attack types but rather on all network traffic and attack scenarios. The features will be required to be small in number to enable a feasible deployment but contain an adequate number of security events to aid in the successful detection of network attacks. The optimal number of dimensions has been identified for all three datasets, which is 20 dimensions. This is indicated in Fig.\ref{vpca2}, where further dimensions gain no additional informational variance. After analysing the DR of each attack type based on the best-performing models, it can be concluded that in a perfect dataset, the number of attack samples needs to be balanced to be efficient in binary classification scenarios.

\section{Conclusions}
In this paper, PCA, autoencoder and LDA have been investigated and evaluated regarding their impact on the classification performance achieved in conjunction with a range of machine learning models. Variance is used to analyse their performance, particularly the correlation between the number of dimensions and detection accuracy. Three deep learning models (DFF, CNN and RNN) and three shallow learning classification algorithms (LR, DT and NB) have been applied to three recent benchmark NIDS datasets, i.e., UNSW-NB15, ToN-IoT and CSE-CIC-IDS2018. In this paper, the optimal combination for each dataset has been mentioned. The optimal number of extracted feature dimensions has been identified for each dataset through an analysis of variance and their impact on the classification performance. However, among the 18 tried combinations of FE algorithm and ML classifiers, no single combination performs best across all three NIDS datasets. Therefore, it is important to note that finding a combination of an FE algorithm and ML classifier that performs well across a wide range of datasets and in practical application scenarios is far from trivial and needs further investigation. While research which aims to improve the intrusion detection and attack classification performance for a particular data and feature set by a few percentage points is valuable, we believe a stronger focus should be placed on the generalisability of the proposed algorithms, especially their performance in more practical network scenarios. In particular, we believe it is crucial to work towards defining generic feature sets that are applicable and efficient across a wide range of NIDS datasets and practical network settings. Such a benchmark feature set would allow a broader comparison of different ML classifiers and would significantly benefit the research community. Finally, explaining the internal operations of ML models would attract the benefits of Explainable AI (XAI) in the NIDS domain.

\bibliography{main.bib}

\end{document}